\documentclass[journal,article,accept,pdftex,moreauthors]{Definitions/mdpi}

\definecolor{purple}{rgb}{0.62, 0.0, 0.77}

\newcommand{\mbr}{\vec{x}}

\usepackage{ulem}

\firstpage{1} 
\pubvolume{1}
\issuenum{1}
\articlenumber{0}
\pubyear{2023}
\copyrightyear{2023}
\datereceived{ } 
\daterevised{ } % Comment out if no revised date
\dateaccepted{ } 
\datepublished{ } 
\hreflink{https://doi.org/} % If needed use \linebreak
\Title{Nuclear matter and finite nuclei: recent studies based on Parity Doublet Model}

\TitleCitation{Nuclear matter and finite nuclei: recent studies based on Parity Doublet Model}

\Author{Yuk-Kei Kong\orcidA{}$^{1,\S,\dagger}$, Youngman Kim$^{2,\ddagger,\dagger}$, and Masayasu Harada$^{3,1,4,\P,\dagger}$}

\AuthorNames{Yuk-Kei Kong, Youngman Kim and Masayasu Harada}
\AuthorCitation{Kong, Y.; Kim, Y.; Harada, M.}
\address{%
$^{1}$ \quad Department of Physics, Nagoya University, Nagoya 464-8602, Japan \\
$^{2}$ \quad Center for Exotic Nuclear Studies, Institute for Basic Science, Daejeon 34126, Korea  \\
$^{3}$ \quad Kobayashi-Maskawa Institute for the Origin of Particles and the Universe, Nagoya University, Nagoya 464-8602, Japan \\
$^{4}$ \quad Advanced Science Research Center, Japan Atomic Energy Agency, Tokai 319-1195, Japan \\
$^\S$ \quad yukkekong2-c@hken.phys.nagoya-u.ac.jp\\
$^\ddag$ \quad ykim@ibs.re.kr\\
$^\P$ \quad harada@hken.pnys.nagoya-u.ac.jp
}

\firstnote{%Current address: Affiliation 3.} 
These authors contributed equally to this work.}
\abstract{
In this review, we summarize recent studies on nuclear matter and finite nuclei based on parity doublet models.
We first construct a parity doublet model (PDM), which includes the chiral invariant mass $m_0$ of nucleons together with the mass generated by the spontaneous chiral symmetry breaking.
We then study the density dependence of the symmetry energy in the PDM, which shows that the symmetry energy is larger for smaller chiral invariant mass.
Then, we investigate some finite nuclei by applying the 
Relativistic Continuum Hartree–Bogoliubov (RCHB) theory to the PDM.
We present the root-mean-square deviation (RMSD) of the binding energies and charge radii, and show that
$m_0 = 700$\,MeV is preferred by the nuclear properties.
Finally, we modify the PDM by adding the iso-vector scalar meson $a_0(980)$ and show that the inclusion of the $a_0(980)$ enlarges the symmetry energy of the infinite nuclear matter.
}


\begin{document}

%%%%%%%%%%%%%%%%%%%%%%%%%%%%%%%%%%%%%%%%%%
\section{Introduction}

Spontaneous chiral symmetry breaking plays an important role in low-energy hadron physics, contributing substantially to the generation of hadron masses and the manifestation of mass differences between chiral partners. In recent decades, there has been a growing focus on investigating the restoration of chiral symmetry in hot and dense matter. Nucleon masses will be changed in such extreme conditions, which provides hints for us towards a further understanding to the mass of hadron and further understanding to the dynamics of the strongly interacting matter.

In the traditional linear sigma model, the entire nucleon mass is generated from the spontaneous chiral symmetry breaking, in which the chiral partner to ordinary nucleon is the nucleon itself. When the chiral symmetry is restored, the nucleon and its chiral partner will be degenerate in mass.
However, increasing evidences from the lattice calculations~\cite{PhysRevD.92.014503,2017JHEP...06..034A} show that, 
with increasing temperature, the mass of negative parity baryon decreases to be degenerate with the mass of positive baryon at the critical temperature.

The Parity Doublet Model (PDM) was proposed in Ref.~\cite{PhysRevD.39.2805} as an extended linear sigma model with parity doubling structure to model the parity doubling of nucleon. In the PDM, the excited nucleon such as $N(1535)$ is regarded as the chiral partner to ordinary nucleon, in which the spontaneous symmetry breaking generates the mass difference between them. By considering the symmetry properties of the chiral partner, the PDM predicts that the masses of the parity partners are degenerate into a finite mass, 
so called, the chiral invariant mass $m_0$, when the chiral symmmetry is restored. 
In addition to the lattice simulations mentioned above, recent analysis based on the QCD sum rules~\cite{PhysRevD.105.014014} also supports the existence of the chiral invariant mass.
Therefore, quantitative and qualitative study of the chiral invariant mass will help us to understand the origin of hadron masses.

Studying the chiral invariant mass $m_0$ is an essential  measure to the origin of the mass of a nucleon. There are several analyses to determine the value of $m_0$ by studying the nucleon properties in vacuum.
For example, the analysis in Ref.~\cite{10.1143/PTP.106.873} shows that $m_0$ is smaller than $500$\,MeV using the decay width of $N(1535)$, while Ref.~\cite{Yamazaki:2018stk} includes higher derivative interaction which makes the large $m_0$ consistent with the decay width.

Chiral symmetry is expected to be partially restored in the high density region, study of which will provide some information on the chiral invariant mass.
Actually, the PDM is applied to study the high density matter in several analyses such as in 
Refs.~\cite{Hatsuda:1988mv, Zschiesche:2006zj, Dexheimer:2007tn, Dexheimer:2008cv, Sasaki:2010bp, Sasaki:2011ff,%
Gallas:2011qp, Paeng:2011hy,%
Steinheimer:2011ea,Dexheimer:2012eu, Paeng:2013xya,Benic:2015pia,Motohiro:2015taa,%
Mukherjee:2016nhb,Suenaga:2017wbb,Takeda:2017mrm,Mukherjee:2017jzi,Paeng:2017qvp,%
Marczenko:2017huu,Abuki:2018ijb,Marczenko:2018jui,Marczenko:2019trv,Yamazaki:2019tuo,%
Harada:2019oaq,Marczenko:2020jma,Harada:2020etl,%
PhysRevC.103.045205,Marczenko:2021uaj,PhysRevC.104.065201,Marczenko:2022hyt,%
PhysRevC.106.065205,Minamikawa:2023eky,Marczenko:2023ohi%
}.
Recently in Refs.~\cite{PhysRevC.103.045205,PhysRevC.104.065201,PhysRevC.106.065205,Minamikawa:2023eky}, the EoS of neutron star (NS) matter constructed from an extended PDM ~\cite{Motohiro:2015taa} is connected to the one from
the NJL-type quark model following Refs.~\cite{Baym_2018,Baym_2019}.
The analysis of Ref.~\cite{PhysRevC.103.045205} used the observational data of NS given in 
Refs.~\cite{NANOGrav:2019jur,LIGOScientific:2017vwq,LIGOScientific:2017ync, TheLIGOScientific:2017qsa,LIGOScientific:2018cki,Miller:2019cac,Riley:2019yda}
to put a constraint on the chiral invariant mass $m_0$ as $600$\,MeV$\lesssim m_0 \lesssim 900$\,MeV, which was updated in Refs.~\cite{PhysRevC.106.065205,Minamikawa:2023eky} to $400$\,MeV$\lesssim m_0 \lesssim 700$\,MeV by considering the effect of anomaly as well as new data 
analysis~\cite{Fonseca:2021wxt,De:2018uhw,Radice:2017lry}.

In recent decades, increasing attention is paid to the effect of isovector-scalar $a_0$(980) meson (or called $\delta$ meson) on asymmetric matter such as NS because it accounts for the attractive force in the iso-vector channel.
References~\cite{Kubis_1997,https://doi.org/10.48550/arxiv.astro-ph/9802303,Miyatsu_2022,Li_2022,https://doi.org/10.48550/arxiv.2209.02861,Thakur_2022,Liu_2005,PhysRevC.80.025806,Gaitanos_2004,PhysRevC.67.015203,PhysRevC.65.045201} use Walecka-type relativistic mean-field (RMF) models, and Refs.~\cite{PhysRevC.90.055801,PhysRevC.84.054309} use density-dependent RMF models to study the effect of $a_0$(980) meson to the symmetry energy as well as to the EoS of asymmetric matter. It was pointed that the existence of $a_0$ meson increases the symmetry energy~\cite{Kubis_1997,Miyatsu_2022,Li_2022,Liu_2005,PhysRevC.80.025806,Gaitanos_2004,PhysRevC.67.015203,PhysRevC.65.045201}, and that it stiffens the NS EoS~\cite{https://doi.org/10.48550/arxiv.astro-ph/9802303,Liu_2005,Li_2022,Miyatsu_2022,Thakur_2022} and asymmetric matter EoS~\cite{PhysRevC.84.054309}. 
Therefore, the $a_0(980)$ meson is influential for the study of asymmetric matter. 
Recently, in Ref.~\cite{PhysRevC.108.055206}, the effect of $a_0(980)$ in neutron star is studied in the PDM and the constraint to the chiral invariant mass is obtained as
$580$\,MeV$\lesssim m_0 \lesssim 860$\,MeV. In particular, this work shows that the $a_0(980)$ meson has large influence to the symmetry energy at density larger than saturation density. Therefore, it is expected that further experimental constraints on the symmetry energy will provide hints to the chiral invariant mass and the origin of the mass of a nucleon.

To put an additional constraint on the value of the chiral invariant mass, 
 the properties of stable nuclei were studied in Ref.~\cite{mun2023} with the PDM in the frame work of a self-consistent relativistic  mean field theory. For the nuclear structure calculations, the Relativistic Continuum Hartree-Bogoliubov (RCHB) theory~\cite{Meng:2005jv} was employed. It was found in Ref.~\cite{mun2023} that  the calculated binding energies and charge radii of
 selected fifteen nuclei are closet to the experimental values  when $m_0=700$ MeV.

In this review, 
we summarize
the recent works on the study of chiral invariant mass in infinite nuclear matter 
in Ref.~\cite{PhysRevC.108.055206}
and finite nuclei 
in Ref.~\cite{mun2023}.
In section~\ref{sec:PDMconstr}, we introduce a PDM based on the chiral SU(2)$_L\times$SU(2)$_R$ symmetry as constructed in Ref.~\cite{Motohiro:2015taa}. Then we construct the infinite nuclear matter using mean field approximation and study the symmetry energy. In section~\ref{sec:PDM FN}, the construction of finite nuclei in mean field model using Relativistic Continuum Hartree–Bogoliubov (RCHB) theory is introduced. After a brief introduction on the construction of finite nuclei, the finite nuclei are constructed using PDM as in Ref.~\cite{mun2023}, and the method to constrain the value of chiral invariant mass using experimental data of finite nuclei are discussed. Some results on the specific nuclei such as the nuclei properties and effective mass of a nucleon in finite nuclei are also shown. In section~\ref{sec:a0}, we review an extension of the PDM by including the isovector scalar meson $a_0(980)$ done in Ref.~\cite{PhysRevC.108.055206}. We also compute the results for the extended PDM without vector meson mixing interaction for comparison. The symmetry energy for these models are compared to the PDM without $a_0$ meson introduced in section~\ref{sec:PDM MF}. Finally, a summary is given in section~\ref{sec:summary}.

%%%%%%%%%%%%%%%%%%%%%%%%%%%%%%%%%%%%%%%%%%
\section{Dense nuclear matter with Parity Doublet Model}
\label{sec:PDM MF}

\subsection{An SU(2)$_L \times $SU(2)$_R$ Parity Doublet Model}\label{sec:PDMconstr}

Here we introduce the parity doublet model (PDM) based on the SU(2)$_L \times$SU(2)$_R$ chiral symmetry constructed in Ref.~\cite{Motohiro:2015taa}. The Lagrangian is given by 
\begin{equation}
    \mathcal{L} = \mathcal{L}_N + \mathcal{L}_M  +  \mathcal{L}_{V}\ , \label{PDM-L1}
\end{equation} 
where $\mathcal{L}_N$ is for the nucleons, $\mathcal{L}_M$ for the scalar and pseudoscalar mesons and $\mathcal{L}_{V}$ for the vector mesons.

In ${\mathcal L}_M$,
% our model, 
the scalar meson field $M$ is introduced as the $(2,2)$ representation under the SU(2)$_L \times $SU(2)$_R$ symmetry which transforms as 
\begin{equation}
    M \rightarrow  g_L M g_R^{\dagger}\ ,
\end{equation} 
where $g_{R,L} \in \mbox{SU}(2)_{R,L} $. %and $g_L \in SU(2)_L$, 
%$M$ carries a $U(1)_A$ charge of -2. $M$ is explicitly 
We parameterize $M$ as 
\begin{equation}
    M = \sigma + i\vec{\pi} \cdot \vec{\tau},
    \label{Mmotohiro}
\end{equation} 
where $\sigma, \vec{\pi}$ are fields for the sigma meson and pions, respectively, 
and $\vec{\tau}$ the Pauli matrices. The vacuum expectation value (VEV) of $M$ is 
\begin{equation}
\begin{aligned}
    \langle 0 | M | 0 \rangle = \begin{pmatrix}
\sigma_0 & 0\\
0 & \sigma_0
\end{pmatrix},
\end{aligned}
\label{def M 2}
\end{equation} 
where $\sigma_0 = \langle 0|\sigma|0 \rangle$ is the VEV of the $\sigma$ field which is equal to the pion decay constant $f_\pi = 93 $ MeV in vacuum.
The explicit form of the Lagrangian $\mathcal{L}_M$ is given by
\begin{equation}
\mathcal{L}_M = \frac{1}{4} \mbox{tr}\left[\partial_\mu M \partial^\mu M^\dagger \right] - V_M\ ,
\end{equation} 
where $V_M$ is the potential for $M$. In the present model, $V_M$ is taken as~\cite{Motohiro:2015taa}
\begin{eqnarray}
V_M&=&-\frac{\bar{\mu}^2}{4}\mbox{tr}[M^\dagger M ]+\frac{\lambda_{4}}{8}\mbox{tr}[(M^\dagger M)^2] -\frac{\lambda_{6}}{12}\mbox{tr}[(M^\dagger M)^3] \nonumber \\ 
 & &-\frac{m^2_{\pi}f_{\pi}}{4}\mbox{tr}[M+M^\dagger]
%-\frac{K}{8}\{detM+detM^\dagger\} 
\ .
\label{VM}
\end{eqnarray} 
In the above potential, the first three terms are invariant under SU(2)$_L\times$SU(2)$_R$
symmetry.
%with the mass dimension {YK,\it it is confusing and seems redundant:less than or equal to four}. 
%The term with the coefficient $\lambda_{6}$ is the six-point interaction term that is introduced in Ref.~\cite{Motohiro:2015taa}
%to reproduce the nuclear saturation properties. 
The fourth term in $V_M$ is  for the explicit chiral symmetry breaking  due to the non-zero current quark masses, which explicitly breaks the SU(2)$_L\times$SU(2)$_R$ symmetry.

For the vector meson, the iso-triplet $\rho$ meson and iso-singlet $\omega$ meson are included based on the hidden local symmetry (HLS)~\cite{PhysRevLett.54.1215,Bando:1987br,Harada:2003jx}
to account for the repulsive force in the hadronic matter. The HLS is introduced by performing polar decomposition of the field $M$ as 
\begin{equation}
\begin{aligned}
M = \xi_L^\dagger S \xi_R \, ,
\end{aligned}
\label{HLSM}
\end{equation}
where $S$ is the scalar meson field. In the case of SU(2)$_L \times $SU(2)$_R$ symmetry with the parameterization in Eq.~(\ref{Mmotohiro}), $S=\sigma$. Moreover, $\xi_{L,R}$ transform as 
\begin{equation}
\begin{aligned}
\xi_{L,R} \rightarrow h_\omega h_\rho \xi_{L,R} g_{L,R}^\dagger ,
\end{aligned}
\label{HLSXItr}
\end{equation}
with $h_\omega \in \mbox{U(1)}_{\rm HLS}$ and $h_\rho \in \mbox{SU(2)}_{\rm HLS}$. In the unitary gauge of the HLS, $\xi_{L,R}$ are parameterized as
\begin{equation}
%\begin{aligned}
\xi_R = \xi_L^\dagger %= exp(i P_a T^a/f_\pi),
= \exp \left( i \pi / f_\pi\right) \ ,
%\end{aligned}
\label{HLSXI}
\end{equation}
where 
%\kongb{
%$\pi$ is the pseudoscalar field. With parameterization Eq.~(\ref{Mmotohiro}), 
%}
$\pi = \sum_{a=1}^{3} \pi^a \tau_a/2$ is the $2\times2$ matrix field for pions with $\tau_a$ being the Pauli matrix.
In the HLS, vector mesons are introduced as the gauge bosons of the HLS. They transform as  
\begin{equation}
\begin{aligned}
\omega_\mu \rightarrow h_\omega \omega_\mu h_\omega^\dagger + \frac{i}{g_\omega}\partial_\mu h_\omega h_\omega^\dagger ,
\end{aligned}
\label{HLSomega}
\end{equation}
\begin{equation}
\begin{aligned}
\rho_\mu \rightarrow h_\rho \rho_\mu h_\rho^\dagger + \frac{i}{g_\rho}\partial_\mu h_\rho h_\rho^\dagger ,
\end{aligned}
\label{HLSrho}
\end{equation}
where $\omega_\mu$ and $\rho_\mu = \sum_{a=1}^3 \rho_\mu^a \tau_a/2$ are the gauge-boson fields for SU(2)$_{\rm HLS}$ and U(1)$_{\rm HLS}$, respectively, and $g_\omega$ and $g_\rho$ are the corresponding gauge coupling constants.

To construct the Lagrangian invariant under the HLS, it is convenient to define the covariantized Maurer-Cartan 1-forms:
\begin{equation}
    \hat{\alpha}_\perp^\mu \equiv \frac{1}{2i}[D^\mu \xi_R \xi_R^\dagger - D^\mu \xi_L \xi_L^\dagger] ,
\end{equation}
\begin{equation}
    \hat{\alpha}_\parallel^\mu \equiv \frac{1}{2i}[D^\mu \xi_R \xi_R^\dagger + D^\mu \xi_L \xi_L^\dagger] ,
\end{equation}
where the covariant derivatives of $\xi_{L,R}$ are given by 
\begin{equation}
   D^\mu \xi_L =  \partial^\mu \xi_L - ig_\rho \rho^\mu \xi_L - ig_\omega \omega^\mu \xi_L  + i \xi_L \mathcal{L}^\mu ,
\end{equation}
\begin{equation}
   D^\mu \xi_R =  \partial^\mu \xi_R - ig_\rho \rho^\mu \xi_R - ig_\omega \omega^\mu \xi_R + i \xi_R \mathcal{R}^\mu .
\end{equation}
Here, $\mathcal{L}^\mu$ and $\mathcal{R}^\mu$ are the external gauge fields corresponding to SU(2)$_L \times$SU(2)$_R$ chiral symmetry. 
Then, the HLS-invariant Lagrangian  including the vector mesons  is given by 
\begin{equation}
\begin{aligned}
\mathcal{L}_{V} {} & =  a_{VNN} \left[ \Bar{N}_{1l} \gamma^\mu \xi_L^\dagger \hat{\alpha}_{\parallel \mu} \xi_L N_{1l} + \Bar{N}_{1r} \gamma^\mu \xi_R^\dagger \hat{\alpha}_{\parallel \mu} \xi_R N_{1r} \right] \\ 
    & \quad + a_{VNN} \left[ \Bar{N}_{2l} \gamma^\mu \xi_R^\dagger \hat{\alpha}_{\parallel \mu} \xi_R N_{2l} + \Bar{N}_{2r} \gamma^\mu \xi_L^\dagger \hat{\alpha}_{\parallel \mu} \xi_L N_{2r} \right] \\ 
    & \quad + a_{0NN} \sum_{i = 1,2} \left[ \Bar{N}_{il} \gamma^\mu \mbox{tr} [ \hat{\alpha}_{\parallel \mu} ] N_{il} + \Bar{N}_{ir} \gamma^\mu  \mbox{tr} [\hat{\alpha}_{\parallel \mu} ] N_{ir} \right] \\
    & \quad + \frac{{m_\rho}^2}{{g_\rho}^2}\mbox{tr} [\hat{\alpha}_\parallel^\mu \hat{\alpha}_{\parallel \mu}]  +  \left(\frac{{m_\omega}^2}{8{g_\omega}^2} - \frac{{m_\rho}^2}{2{g_\rho}^2}\right) \mbox{tr} [\hat{\alpha}_\parallel^\mu]\mbox{tr}[ \hat{\alpha}_{\parallel \mu}] \\
    &\quad - \frac{1}{8{g_\omega}^2}\mbox{tr}[\omega^{{\mu\nu}}\omega_{{\mu\nu}}] - \frac{1}{2{g_\rho}^2}\mbox{tr}[\rho^{{\mu\nu}}\rho_{{\mu\nu}}],
\
\end{aligned}
\label{Lv}
\end{equation}
where $\rho^{{\mu\nu}}$ and $\omega^{{\mu\nu}}$ are the field strengths of the $\rho$ meson and the $\omega$ meson given by
\begin{align}
\rho_{\mu\nu} = & \partial_\mu \rho_\nu -  \partial_\nu \rho_\mu - i g_\rho \, \left[ \rho_\mu \,,\, \rho_\nu \right] \ , \notag\\
\omega_{\mu\nu} = & \partial_\mu \omega_\nu -  \partial_\nu \omega_\mu \ .
\end{align}

Finally, the baryonic Lagrangian $\mathcal{L}_N$ based on the parity doubling structure \cite{PhysRevD.39.2805,10.1143/PTP.106.873} is given by 
\begin{equation}
\begin{aligned}
\mathcal{L_N} {} & = \Bar{N_{1}}i \gamma^\mu \mathcal{D}_\mu N_{1} + \Bar{N_{2}}i \gamma^\mu \mathcal{D}_\mu N_{2}  \\
 & \quad - m_0 [\Bar{N}_{1}\gamma_5 N_{2} - \Bar{N}_{2}\gamma_5 N_{1}] \\
      & \quad - g_1 [\Bar{N}_{1l} M N_{1r} + \Bar{N}_{1r} M^{\dagger}  N_{1l}]\\
       & \quad - g_2 [\Bar{N}_{2r} M N_{2l} + \Bar{N}_{2l} M^{\dagger}  N_{2r}], \
\end{aligned}
\label{eq231}
\end{equation}
where $N_{ir} = \frac{1+\gamma_5}{2} N_i$ ($N_{il}= \frac{1-\gamma_5}{2}N_i$) ($i=1,2$) is the right-handed (left-handed) component of the nucleon fields $N_i$ and the covariant derivatives of the nucleon fields are defined as 
\begin{equation}
    \begin{aligned}
        \mathcal{D}^\mu N_{1l,2r} & = \left( \partial^\mu - i \mathcal{L}^\mu - i {\mathcal V}^\mu \right) N_{1l,2r}\ ,\\
        \mathcal{D}^\mu N_{1r,2l} & = \left( \partial^\mu - i \mathcal{R}^\mu - i {\mathcal V}^\mu \right) N_{1r,2l}\ ,
    \end{aligned}
\label{eq231}
\end{equation}
where ${\mathcal V}^\mu$ is the external gauge field corresponding to the U(1) baryon number.
By diagonalizing $\mathcal{L}_N$, we obtain two baryon fields $N_+$ and $N_-$ corresponding to the positive parity and negative parity nucleon fields, respectively.
Their masses in vacuum are obtained as~\cite{PhysRevD.39.2805,10.1143/PTP.106.873}
\begin{equation}
\begin{aligned}
    m^{\rm(vac)}_{\pm} = \frac{1}{2} \bigg[ \sqrt{(g_1+g_2)^2\sigma_0^2 + 4m_0^2} \pm (g_1 - g_2)\sigma_0 \bigg]\ .
\end{aligned}
\label{mvaj}
\end{equation} 
%A detailed review of this diagonalization procedure is available in \cite{10.1143/PTP.106.873}. 
In the present work, $N_+$ and $N_-$ are identified as $N$(939) and $N(1535)$, respectively.

\subsection{Dense nuclear matter in PDM with mean field approximation}\label{sec:PDMmatter}

To construct the nuclear matter from the model introduced in the previous section, we adopt the mean-field approximation following Ref.~\cite{Motohiro:2015taa}, 
%\sout{ \cite{PhysRevC.103.045205,PhysRevC.106.065205}, } 
by taking 
\begin{equation}
    \sigma(x) \rightarrow \sigma, \qquad \pi(x) \rightarrow 0.
\end{equation} 
Then, the mean field for $M$ becomes 
\begin{equation}
\begin{aligned}
    \langle M \rangle = \begin{pmatrix}
\sigma   & 0\\
0 & \sigma
\end{pmatrix}
\ .
\end{aligned}
\end{equation} 
Now, the potential $V_M$ is written in terms of the meson mean fields as
\begin{equation}
\begin{aligned}
V_M = &  - \frac{\bar{\mu}^2_{\sigma}}{2} \sigma^2  +  \frac{\lambda_4}{4}  \sigma^4 - \frac{\lambda_6}{6}  \sigma^6  - m^2_{\pi} f_{\pi} \sigma. \
\end{aligned}
\label{VMma}
\end{equation}

In the mean-field approximation, the vector meson fields are taken as
\begin{equation}
    \omega_{\mu} (x) \rightarrow \omega \delta_{\mu 0}, \qquad   \rho^{i}_{\mu} (x) \rightarrow \rho \delta_{\mu 0} \delta_{i 3},
\end{equation}
according to the rotational symmetry and isospin symmetry. 
Subsequently, the Lagrangian of the vector mesons is expressed in terms of the mean fields as
\begin{equation}
\begin{aligned}
\mathcal{L}_V = - g_{\omega NN} \sum_{\alpha j} \Bar{N}_{\alpha j} \gamma^0 \omega N_{\alpha j} - g_{\rho NN} \sum_{\alpha j} \Bar{N}_{\alpha j} \gamma^0  \frac{\tau_3}{2} \rho N_{\alpha j} + \frac{1}{2} m^2_{\omega} \omega^2 + \frac{1}{2} m^2_{\rho} \rho^2.  \
\end{aligned}
\end{equation} 
with 
\begin{align}
    g_{\omega NN} & =  \left(  a_{VNN} + a_{0NN}  \right) g_\omega \ ,\\
    g_{\rho NN} & =  a_{VNN}g_\rho \ .
\end{align}

Then, the thermodynamic potential for the nucleons is written as \begin{equation}
\begin{aligned}
{} & \Omega_{N}   = - 2 \sum_{\alpha=\pm,  j=\pm} \int^{k_f} \frac{d^3p}{(2 \pi)^3} \bigg[ \mu^*_j - \omega_{\alpha j}  \bigg],
\end{aligned}
\label{OFG}
\end{equation} 
where 
$\alpha = \pm$ denotes the parity and $j = \pm $ the iso-spin of nucleons.
$\mu^*_j$ is the effective chemical potential given by \begin{equation}
\begin{aligned}
\mu^*_j \equiv (\mu_B - g_{\omega NN} \omega) + \frac{j}{2} (\mu_I - g_{\rho NN} \rho)\ ,
\end{aligned}
\label{mustar}
\end{equation} 
and $\omega_{\alpha j}$ is the energy of the nucleon
% with parity ($\alpha = \pm 1$) and the isospin ($j = p,n$) 
defined as $\omega_{\alpha j} = \sqrt{(\vec{p})^2 + (m^*_{\alpha j})^2}$ where $\vec{p}$ and $m^*_{\alpha j}$ are the momentum and the effective mass of the nucleon. The effective mass $m^*_{\alpha j}$ is given by 
\begin{equation}
\begin{aligned}
    m^*_{\alpha j} =  \frac{1}{2} \bigg[ \sqrt{(g_1+g_2)^2\sigma^2 + 4m_0^2} + \alpha(g_1 - g_2)\sigma \bigg]\ .
\end{aligned}
\label{maj}
\end{equation}

The entire thermodynamic potential for hadronic matter 
is expressed as 
\begin{equation}
\begin{aligned}
{} & \Omega_H   = \Omega_{N} - \frac{\bar{\mu}^2_{\sigma}}{2} \sigma^2  +  \frac{\lambda_4}{4} \sigma^4 - \frac{\lambda_6}{6} \sigma^6  - m^2_{\pi} f_{\pi} \sigma  - \frac{1}{2} m^2_{\omega} \omega^2 - \frac{1}{2} m^2_{\rho} \rho^2 - \Omega_{0}\ ,
\end{aligned}
\label{eq36m}
\end{equation} 
where we subtracted the potential at the vacuum \begin{equation}
\begin{aligned}
\Omega_0 \equiv - \frac{\bar{\mu}^2_{\sigma}}{2} f_{\pi}^2  +  \frac{\lambda_4}{4}  f_{\pi}^4  
 - \frac{\lambda_6}{6} f_{\pi}^6  - m^2_{\pi} f_{\pi}^2\ .
\end{aligned}
\label{eq37}
\end{equation}

%%%%%%%%%%%%%%%%%%%%%%%%%%%%%%%%%%%%%%%%%%
\subsection{Nuclear saturation properties}\label{sec:NSP}

Nuclear properties at the saturation density $n_0$ = 0.16 $fm^{-3}$ are very important to be satisfied in nuclear physics. At the saturation, the energy per nucleon of the infinite symmetric nuclear matter is minimized. There are several fundamental nuclear properties at the saturation density: the binding energy $B_0$, the nuclear incompressibility $K_0$, the nuclear symmetry energy $S_0$, and the slope parameter $L_0$. In the present work, the model parameters are determined such that the saturation properties of the nuclear matter are reproduced.

We first obtain the pressure of hadronic matter $P$ from the thermodynamic potential in Eq.~(\ref{eq36m}) as
\begin{equation}
P(\mu_B, \mu_I) = - \Omega_H(\mu_B,\mu_H;\sigma=\sigma_0,\omega=\omega_0,\rho=\rho_0) \ ,
\end{equation}
where $\mu_B$ and $\mu_I$ are the chemical potentials for the baryon number and the isospin number, and $\sigma_0$, $\omega_0$ and $\rho_0$ are the solutions of the stationary conditions of $\Omega_H$ given by
\begin{equation}
\frac{\partial \Omega_H}{\partial \sigma} = 0 \ , \quad \frac{\partial \Omega_H}{\partial \omega} = 0 \ , \quad  \frac{\partial \Omega_H}{\partial \rho} = 0 \ .
\end{equation}
From the pressure $P$, we define the baryon number density $n_B$ and the isospin density $n_I$ as
\begin{equation}
n_B = \frac{\partial P}{\partial \mu_B} \ , \quad n_I = \frac{\partial P}{\partial \mu_I} \ .
\end{equation}
They are related to the proton number density $n_p$ and the neutron number denisty $n_n$ as
\begin{equation}
n_B = n_p + n_n \ , \quad n_I = \frac{1}{2}n_p - \frac{1}{2}n_n \ .
\end{equation}
As usual, from these densities and the pressure, 
we obtain the energy density $\epsilon$ via the Legendre transformation as 
\begin{equation}
\epsilon (n_B,n_I) = -P + \mu_Bn_B +  \mu_In_I \ .
\label{eq27}
\end{equation}
It is convenient to define the energy per nucleon as
\begin{equation}
w(x,\delta) \equiv \frac{\epsilon (n_B,n_I)}{n_B} - m_N, 
\label{eq24}
\end{equation}
where
\begin{equation}
x \equiv \frac{n_B - n_0}{3 n_0} \ , \quad \delta \equiv - \frac{2n_I}{n_B} \ .
\end{equation}

At the saturation density $n_B=n_0$, the symmetric nuclear matter ($n_I=0$) forms the most stable state with minimized energy. In other words, $w(x,\delta)$ is stationary when $(x,\delta)=(0,0)$, with $w(0,0) < 0$. Then,
\begin{equation}
\begin{aligned}
\frac{\partial w}{\partial \delta} \Bigr|_{\substack{0}}   {} 
& = \frac{\partial w}{\partial n_B} \frac{\partial n_B}{\partial \delta} + \frac{\partial w}{\partial n_I} \frac{\partial n_I}{\partial \delta} \Bigr|_{\substack{0}} = - \frac{1}{2} \mu_I \Bigr|_{\substack{0}} = 0, \
\end{aligned}
\label{eq25}
\end{equation}

\begin{equation}
\begin{aligned}
\frac{\partial w}{\partial x} \Bigr|_{\substack{0}} {} & = \frac{3P}{n_0} \Bigr|_{\substack{0}} =0 \ ,
\end{aligned}
\label{eq26}
\end{equation}
where  $ \quad \Bigr|_{\substack{0}}$ means that the derivatives are evaluated at $(x,\delta)=(0,0)$. These imply that the pressure $P$ and isospin chemical potential $\mu_I$ are zero at the saturation density.
%\begin{equation}
%\begin{aligned}
%w(0,0) {} & = - \frac{E_B}{A} \Bigr|_{\substack{A\rightarrow \infty \\ N = Z}} \\
%& = - \frac{1}{A} \bigg \{  a_v A + a_s A^{\frac{2}{3}} - a_c \frac{Z(Z-1)}{A^{\frac{1}{3}}}  - a_A \frac{ (N-Z)^2}{A} + a_p A^{-\frac{1}{2}}  \bigg \} \Bigr|_{\substack{A\rightarrow \infty \\ N = Z}}\\
%&  = - a_v \\
%& \approx - 16 \ \text{MeV}. \
%\end{aligned}
%\label{eq29}
%\end{equation}
The binding energy $B_0$ is obtained as 
\begin{equation}
B_0 {}  = - w(0,0)  = - \frac{\epsilon}{n_B} \Bigr|_{\substack{0}} + m_N = - \mu_B \Bigr|_{\substack{0}} +m_N \ .
\label{eq51}
\end{equation}
%\begin{equation}
%\begin{aligned}
%B_0 {} & = - w(0,0) \\
%& = - \frac{\epsilon}{n_B} \Bigr|_{\substack{0}} + m_N \\
%& = - \mu_B \Bigr|_{\substack{0}} +m_N \\
%& = 16 \ \text{MeV}, \
%\end{aligned}
%\label{eq51}
%\end{equation}
%
%\begin{equation}
%\mu_B \Bigr|_{\substack{0}} = - B_0 + m_N = 923 \ \text{MeV}. 
%\label{eq52}
%\end{equation}
In this review, we take $B_0 = -16\,$MeV as an input.

Expanding $w(x,\delta)$ around the stationary point $(x,\delta)=(0,0)$, we obtain
\begin{equation}
\begin{aligned}
w(x, \delta) {} & =  w(0,0) + \frac{1}{2} \frac{\partial^2 w}{\partial x^2} \Bigr|_{\substack{0}} x^2 + \frac{1}{2} \frac{\partial^2 w}{\partial \delta^2} \Bigr|_{\substack{0}} \delta^2  + \frac{1}{2} \frac{\partial^3 w}{\partial x \partial \delta^2} \Bigr|_{\substack{0}} x \delta^2 + O(x^3) \\
& \equiv -B_0 + \frac{1}{2} K_0 x^2 + (S_0 + L_0 x ) \delta^2 + O(x^3) \ ,
\end{aligned}
\label{eq30}
\end{equation}
where $K_0$, $S_0$ and $L_0$ are called as the incompressibility, the symmetry energy and the slope parameter at the saturation density, respectively.

The incompressibility $K_0$ represents the curvature of $w(x,\delta)$ in the direction of the baryon number density. It corresponds to the rate of increase of the baryon chemical potential $\mu_B$ with respect to $n_B$ around the saturation density. $K_0$ is calculated as 
\begin{equation}
K_0 {} \equiv \frac{\partial^2 w}{\partial x^2} \Bigr|_{\substack{0}}  = 9 n_0^2 \frac{\partial^2}{\partial n_B^2} \bigg( \frac{\epsilon}{n_B} \bigg) \Bigr|_{\substack{0}}  = 9 n_0 \frac{\partial \mu_B}{\partial n_B} \Bigr|_{\substack{0}}.
\label{eq58}
\end{equation}
%
%
%
%\begin{equation}
%\begin{aligned}
%K_0 {} & \equiv \frac{\partial^2 w}{\partial x^2} \Bigr|_{\substack{0}} \\
%&  = 9 n_0^2 \frac{\partial^2}{\partial n_B^2} \bigg( \frac{\epsilon}{n_B} \bigg) \Bigr|_{\substack{0}} \\
%& = 9 n_0 \frac{\partial \mu_B}{\partial n_B} \Bigr|_{\substack{0}}.
%\end{aligned}
%\label{eq58}
%\end{equation}
We note that $K_0$ corresponds to the hardness of the (symmetric) matter around the saturation density; the larger $K_0$ corresponds to the larger pressure at high baryon density. Thus, it is called the incompressibility of nuclear matter because a larger $K_0$ corresponds to a matter that is more resistant to compression. The generally accepted values are $K_0 = 240 \pm 40$ (see recent review \cite{GARG201855} for detailed discussion and summary of the values of $K_0$.) In this review, the results with $K_0$ = 215, 240 MeV are computed for comparison.

The symmetry energy $S_0$ is defined to be the slope of $w(x,\delta)$ in the isospin density direction around $n_0$ as
\begin{equation}
%\begin{aligned}
S_0 {}  \equiv \frac{1}{2} \frac{\partial^2 w}{\partial \delta^2} \Bigr|_{\substack{0}}  =  \frac{n_0^2}{8} \frac{\partial^2}{\partial n_I^2} \bigg( \frac{\epsilon}{n_B} \bigg) \Bigr|_{\substack{0}}  = \frac{n_0}{8}  \frac{\partial \mu_I}{\partial n_I} \Bigr|_{\substack{0}}.
%\end{aligned}
\end{equation}
The symmetry energy is the energy that arises from the asymmetry of the matter. If we ignore $O(x^3)$ contribution in Eq.~(\ref{eq30}), the symmetry energy at the saturation density $S_0$ can be approximated by
\begin{equation}
    S_0 \approx w(0,1)-w(0,0),
\end{equation}
which is the energy difference between pure neutron matter and symmetric matter. Then, the term $ S_0 \delta^2$ can be seen as the energy arises from the difference of $n_p$ and $n_n$ (the asymmetry of the matter) around the saturation density. For later convenience, we define the symmetry energy at arbitrary baryon density $n_B$ as
\begin{equation}
    S(n_B) \equiv \frac{1}{2} \frac{\partial^2 w(x,\delta)}{\partial \delta^2} \Bigg\vert_{\delta = 0 } \ .
    \label{Sbeq}
\end{equation}
This $S(n_B)$ approximately corresponds to the energy difference between pure neutron matter and symmetric matter at $n_B $:
\begin{equation}
    S(n_B) \approx w(x,1)-w(x,0).
    \label{Srhob}
\end{equation}
The value of $S_0$ is well-studied with little ambiguity. In this review, $S_0$ is taken to be 31 MeV.

Finally, the slope parameter $L_0$ is given by
\begin{equation}
\begin{aligned}
L_0 {} & \equiv \frac{1}{2} \frac{\partial^3 w}{\partial x \partial \delta^2} \Bigr|_{\substack{0}}  %\\
%&
=\frac{\partial S(n_B)}{\partial x}      \Bigr|_{\substack{0}} %\\
%& 
=  3 S_0 + \frac{3 n_0^2}{8}  \frac{\partial^2 \mu_I}{\partial n_B n_I} \Bigr|_{\substack{0}}.
\end{aligned}
\end{equation}
The slope parameter approximates the slope of the symmetry energy in the direction of baryon number density around the saturation density. The larger $L_0$ results in the larger symmetry energy $S(n_B)$ at higher density. Due to the experimental difficulties, the value of $L_0$ possesses large uncertainty and has been discussed for many years. The recent accepted values are $L_0 = 57.7\pm 19$ MeV as summarized in Ref.~\cite{universe7060182}.
\begin{table}[h] 
\caption{Saturation properties that are used to determine the model parameters: saturation density $n_0$, binding energy $B_0$, incompressibility $K_0$, and symmetry energy $S_0$.\label{SP}}
\newcolumntype{C}{>{\centering\arraybackslash}X}
\begin{tabularx}{\textwidth}{CCCCC}
\toprule
\textbf{$n_0$[$fm^{-3}$]}&\textbf{$B_0$[MeV]}&\textbf{$K_0$[MeV]}&\textbf{$S_0$[MeV]}\\
\midrule
0.16 & 16 & 215, 240 & 31 \\
\bottomrule
\end{tabularx}
\end{table}

\subsection{Determination of model parameters}

In the present model, the model parameters are fitted to reproduce the nuclear saturation properties as well as physical masses and
the decay constant in vacuum. There are seven parameters to be determined for a given value of the chiral invariant mass $m_0$: 
\begin{equation}
    g_1\,,\ g_2\,,\ \bar{\mu}^2_{\sigma}\,, \ \lambda_4\, ,\ \lambda_6\,,\ g_{\omega NN}\,,\   g_{\rho NN}\,.
\end{equation}
The vacuum expectation value of $\sigma$ is taken to be $\sigma_0=f_{\pi}$ with the pion decay constant $f_{\pi}$=93 MeV. The Yukawa coupling constants
$g_1$ and $g_2$ are determined by fitting them to 
the nucleon masses in vacuum given in Eq.~(\ref{mvaj}), with $m_+ = m_{N}=939$\,MeV and $m_{-} = m_{N^*}=1535$\,MeV for fixed value of the chiral invariant mass $m_0$. The values of 
$\bar{\mu}^2_{\sigma}$, $\lambda_4$, $\lambda_6$, $g_{\omega NN}$, and $g_{\rho NN}$ are determined by the saturation properties shown in Table~\ref{SP} together with the stationary condition of the potential in vacuum given by
\begin{equation}
%\bar{\mu}_\sigma^2 \sigma + \lambda_4 \sigma^3 - \lambda_6 \sigma^5 - m_\pi^2 f_\pi = 0 \ .
\bar{\mu}_\sigma^2 f_\pi - \lambda_4 f_\pi^3 + \lambda_6 f_\pi^5 + m_\pi^2 f_\pi = 0 \ .
\label{vacuum condition}
\end{equation}
For the meson masses, we use the values listed in Table~\ref{vacM}.
\begin{table}[htbp] 
\caption{Values of meson masses in vacuum and pion decat constant in unit of MeV.\label{vacM}}
\newcolumntype{C}{>{\centering\arraybackslash}X}
\begin{tabularx}{\textwidth}{CCCC}
\toprule
\textbf{$m_\pi$}&
\textbf{$m_\omega$}&
\textbf{$m_\rho$}&
\textbf{$f_\pi$}\\
\midrule
140 & 783 & 776 & 93\\
\bottomrule
\end{tabularx}
\end{table}
We should note that there is a relatively large uncertainty in the incompressibility, so that we use $K_0=215$ and $240$\,MeV as inputs for studying the dependence. The determined values of the parameters for a fixed value of $m_0$ are summarized in Table~\ref{PD1215}.
%%%%%%%%%%%%%%%%%%%%%%%%%%
\begin{table}[H] 
\caption{Values of $g_1$, $g_2$, $\bar{\mu}^2_{\sigma},  \lambda_4, \lambda_6, g_{\omega NN}, g_{\rho NN}$ for $m_0$ = $600$-$900$ MeV, $K_0 = 215, 240$\,MeV.}
\label{PD1215}
\newcolumntype{C}{>{\centering\arraybackslash}X}
\begin{tabularx}{\textwidth}{CCCCCC}
\toprule
 & \textbf{$m_0$ (MeV)}	& \textbf{600}	& \textbf{700}  & \textbf{800}  & \textbf{900}\\
\midrule
\multirow[m]{7}{*}{$K_0 = 215$ MeV} & $g_1$&8.427 & 7.762 & 6.941 & 5.921\\
& $g_2$&14.836 & 14.171 & 13.349 & 12.329\\
& $\bar{\mu}^2_{\sigma}/f^2_{\pi}$&23.377 & 20.979 & 13.346 & 2.502\\
& {{$\lambda_4$}}&42.368 & 38.92 & 26.128 & 6.673\\
& {{$\lambda_6 f^2_{\pi}$}}&16.79 & 15.739 & 10.58 & 1.969\\
& {{$g_{\omega NN}$}}&8.902 & 7.055 & 5.471 & 3.389\\
& {{$g_{\rho NN}$}}&7.896 & 8.16 & 8.314 & 8.442 \\
%& {{$g_{\rho NN}$}}&3.948 & 4.08 & 4.157 & 4.221 \\
\midrule
\multirow[m]{7}{*}{$K_0 = 240$ MeV} & $g_1$&8.427 & 7.762 & 6.941 & 5.921\\
& $g_2$&14.836 & 14.171 & 13.349 & 12.329 \\
& $\bar{\mu}^2_{\sigma}/f^2_{\pi}$& 21.821 & 18.842 & 11.692 & 1.537\\
& {{$\lambda_4$}}&39.367 & 34.583 & 22.577 & 4.388 \\
& {{$\lambda_6 f^2_{\pi}$}}&15.344 & 13.54 & 8.683 & 0.649\\
& {{$g_{\omega NN}$}}&9.132 & 7.305 & 5.66 & 3.522 \\
& {{$g_{\rho NN}$}}&7.854 & 8.13 & 8.298 & 8.436\\
%& {{$g_{\rho NN}$}}&3.927 & 4.065 & 4.149 & 4.218\\
\bottomrule
\end{tabularx}
\end{table}

In the present model, the slope parameter $L_0$ is computed as an output. The resultant values are shown in Table~\ref{mL0}. We note that the computed $L_0$ is only slightly larger than the recently accepted values $L_0 = 57.7\pm 19$ MeV as summarized in Ref.~\cite{universe7060182}.
\begin{table}[H] 
\caption{Slope parameter $L_0$ computed as a output from the model.}
\label{mL0}
\newcolumntype{C}{>{\centering\arraybackslash}X}
\begin{tabularx}{\textwidth}{CCCCCC}
\toprule
 & \textbf{$m_0$ (MeV)}	& \textbf{600}	& \textbf{700}  & \textbf{800}  & \textbf{900}\\
\midrule
\multirow[m]{1}{*}{$K_0 = 215$\,MeV} & $L_0$&85.91 & 82.87 & 81.32 & 80.15\\
\midrule
\multirow[m]{1}{*}{$K_0 = 240$\,MeV} & $L_0$&86.25 & 83.04 & 81.33 & 80.08\\
\bottomrule
\end{tabularx}
\end{table}

\subsection{symmetry energy}
\label{Smoto}

In this section, we study the symmetry energy as defined in Eq.~(\ref{Sbeq}). In the present model, the symmetry energy at arbitrary baryon density is given by
\begin{equation}
%\begin{aligned}
S(n_B)  = \frac{n_B}{8} 
    \frac{\partial \mu_I}{\partial n_I} \Bigr|_{\substack{n_I=0}}
=  \frac{(k^*_+)^2}{6 \mu^*_+}   + \frac{n_B}{2}\frac{(g_{\rho NN}/2)^2}{m_{\rho}^2 } \ .
    \label{SrhoB}
%\end{aligned}
\end{equation} 
where $\mu^*_+ \equiv \mu^*_p \big|_{\substack{n_I=0}}=\mu^*_n \big|_{\substack{n_I=0}}$ is the effective chemical potential for $N(939)$ in the symmetric matter, $k^*_{+} \equiv \sqrt{(\mu^*_p)^2 - (m^*_{+p})^2}  \big|_{\substack{n_I=0}} = \sqrt{(\mu^*_n)^2 - (m^*_{+n})^2}  \big|_{\substack{n_I=0}}$ the corresponding Fermi momentum and $m^*_{+} \equiv m^*_{+p} \big|_{\substack{n_I=0}} = m^*_{+n} \big|_{\substack{n_I=0}}$ the effective mass.

\begin{figure}[H]
\centering
\includegraphics[width=10.5 cm]{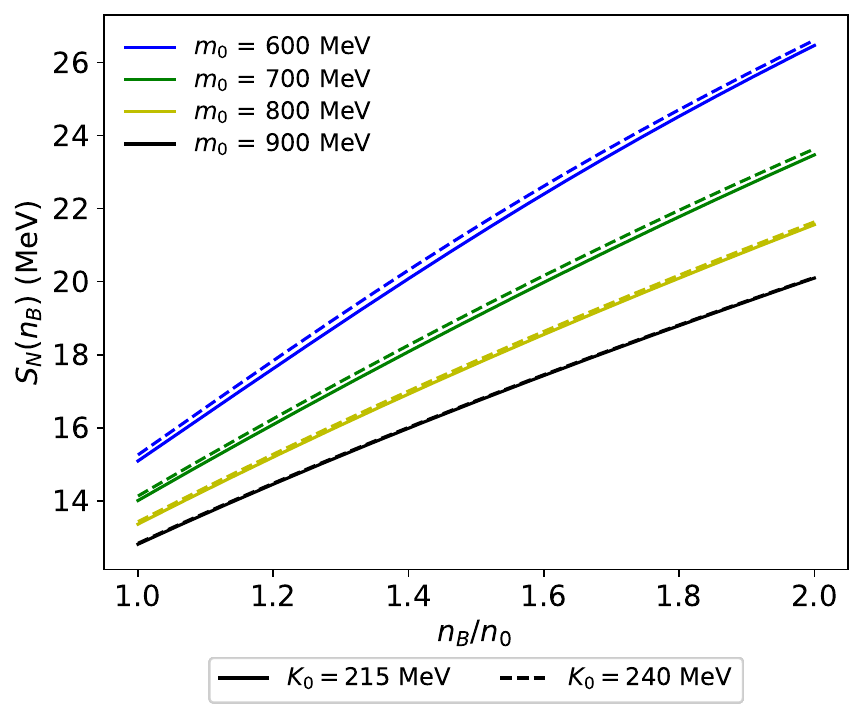}
\caption{Nucleon contribution $S_N(n_B)$ for $m_0 = 600$-$900$\,MeV. Solid curves represent $S_N(n_B)$ with $K_0 = 215$\,MeV, while dashed curves represent $S_N(n_B)$ with $K_0 = 240$\,MeV. \label{SbN without VM}}
\end{figure}

The symmetry energy is divided into two contributions: the nucleon contribution and the $\rho$ meson contribution. The nucleon contribution $S_N(n_B)$ is given by the expression
\begin{equation}
\begin{aligned}
    S_N(n_B) \equiv  \frac{(k^*_+)^2}{6 \mu^*_+} \ ,
    \label{SrhoN}
\end{aligned}
\end{equation} 
which arises from the effective kinetic contribution of nucleons. Figure~\ref{SbN without VM} shows $S_N(n_B)$ for $m_0 = 600$-$900$\,MeV with $K_0 = 215$, $240$\,MeV. It is observed that $S_N(n_B)$ increases with density, as the effective kinetic energy of nucleons rises with density. Additionally, it is noted that $S_N(n_B)$ is larger for smaller $m_0$ due to the stiffening of matter for smaller $m_0$. It can be also seen that $S_N(n_B)$ is larger for larger $K_0$. However, the change of $K_0$ has little effect on $S_N(n_B)$.

\begin{figure}[H]
\centering
\includegraphics[width=10.5 cm]{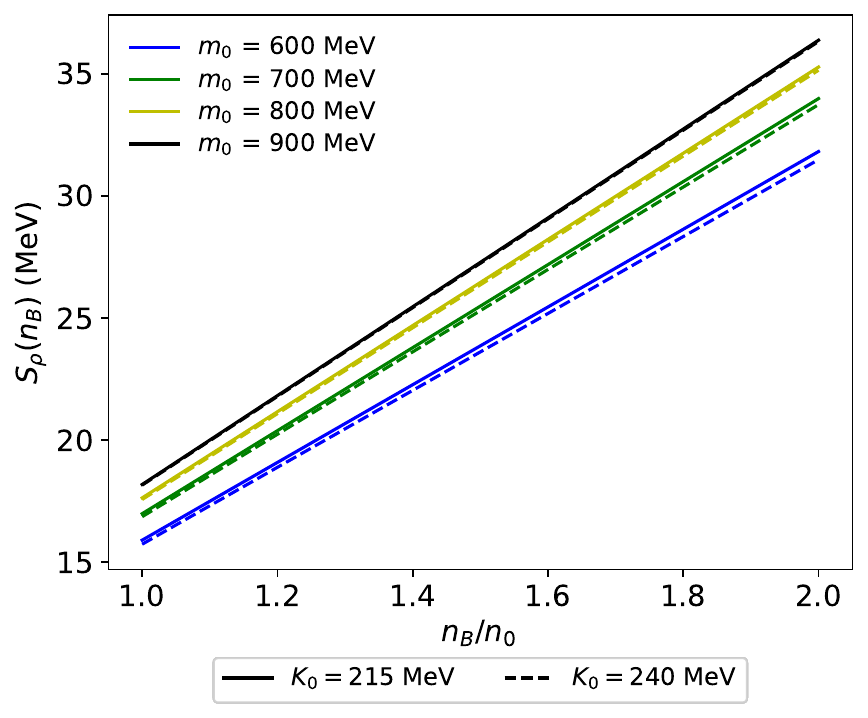}
\caption{$\rho$ meson contribution $S_\rho(n_B)$ for $m_0 = 600$-$900$\,MeV. Solid curves represent $S_\rho(n_B)$ with $K_0 = 215$\,MeV, while dashed curves represent $S_\rho(n_B)$ with $K_0 = 240$\,MeV. \label{Sbrho without VM}}
\end{figure}

Another contribution to the symmetry energy arises from the repulsive interaction of the $\rho$ meson, as described by the expression:
\begin{equation}
\begin{aligned}
    S_\rho(n_B)\equiv \frac{n_B}{2} \frac{(g_{\rho NN}/2)^2}{m_{\rho}^2} \ .
    \label{Srho}
\end{aligned}
\end{equation} 
This shows that the contribution is always positive and thus provides repulsive force to the matter. Figure~\ref{Sbrho without VM} shows the behavior of $S_\rho(n_B)$ for $m_0 = 600$-$900$\,MeV with $K_0 = 215$, $240$\,MeV. It is noteworthy that $S_\rho(n_B)$ is directly proportional to the baryon density $n_B$, rendering it an increasing function with density. We also note that $S_\rho(n_B)$ exhibits larger values for heavier $m_0$. This is understood as follows: at the saturation density, the symmetry energy $S_0$ is fixed to be $31$\,MeV. Since the total symmetry energy is given by Eq.~(\ref{SrhoB}), a larger $m_0$ corresponds to a smaller $S_N(n_0)$ and, consequently, a larger $S_\rho(n_0)$. This larger $S_\rho(n_0)$ yields a larger coupling constant $g_{\rho NN}$ for larger $m_0$. As a result, $S_\rho(n_B)$ is larger for larger $m_0$ at density higher than the saturation density. Figure~\ref{Sbrho without VM} also shows that $K_0$ has little effect on $S_\rho$ similarly to the case for $S_N$.
%\unskip

\begin{figure}[H]
\centering
\includegraphics[width=10.5 cm]{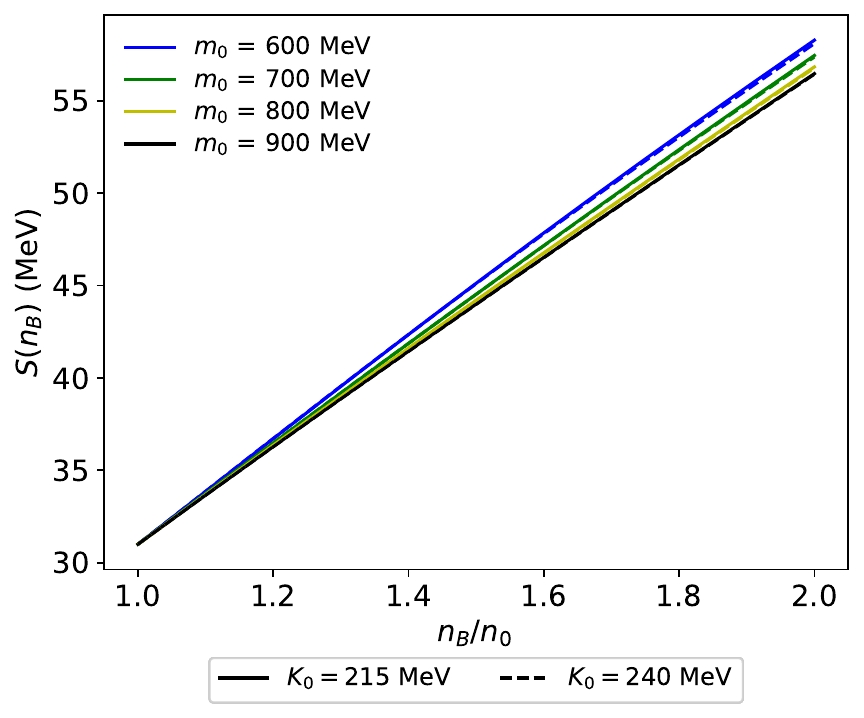}
\caption{Symmetry energy $S(n_B)$ as a function of the baryon number density for $m_0 = 600$-$900$\,MeV. Solid curves represent $S_N(n_B)$ with $K_0 = 215$\,MeV, while dashed curves represent $S_N(n_B)$ with $K_0 = 240$\,MeV.\label{Sb without VM}}
\end{figure}

Figure~\ref{Sb without VM} shows the symmetry energy $S(n_B)$ for $m_0=600$-$900 $\,MeV and $K_0= 215$, $240$\,MeV. We note that the symmetry energy is increasing as the density increases. We also note that the influence of $m_0$ on $S(n_B)$ is relatively small; when increasing $m_0$ from $600$\,MeV to $900$\,MeV, the reduction in $S(2n_0)$ is approximately 3$\%$ for $K_0 = 215$\,MeV. Similarly, the impact of $K_0$ on $S(n_B)$ is even smaller, leading to a change of less than 0.3$\%$ in $S(2n_0)$ for $m_0 = 600$\,MeV.
%\unskip

%%%%%%%%%%%%%%%%%%%%%%%%%%%%%%%%%%%%%%%%%%
\section{Finite nuclei}
\label{sec:PDM FN}

\subsection{Relativistic density functional theory for finite nuclei}
Here, we describe how one obtains the nuclear energy density functional based on the relativistic mean field theory and the corresponding equation of motion for nucleons and mesons. We also discuss in brief how to solve the equation of motion
especially for exotic nuclei in which the continuum effect is important.
The Relativistic Continuum Hartree–Bogoliubov (RCHB) theory~\cite{Meng:2005jv}  is an extension of the relativistic mean field theory in a self-consistent way with both bound and (discretized) continuum states.

The starting Lagrangian is given by 
\begin{eqnarray}
{\mathcal L}\!\!\!\! &=&\!\!\!\!\bar\psi [i\!\!\not\!\partial-M-g_\sigma \sigma-g_\omega \!\!\not\!\omega 
-g_\rho \!\!\not\!\rho -e\!\!\not\!\! A \frac{1-\tau_3}{2}]\psi 
 + \frac{1}{2} \partial^\mu\sigma\partial_\mu\sigma-U(\sigma)\nonumber \\
&&-\Omega^{\mu\nu}\Omega_{\mu\nu}+U_\omega(\omega_\mu)
-\frac{1}{4}\vec{R}^{\mu\nu}\cdot \vec{R}_{\mu\nu}+U_\rho(\vec{\rho}_\mu)
-\frac{1}{4}F_{\mu\nu}F_{\mu\nu}\, ,
\end{eqnarray}
where $\Omega^{\mu\nu}$, $\vec{R}^{\mu\nu}$ and $F_{\mu\nu}$ are the field strength tensors of the $\omega$ meson, $\rho$ meson
and electromagnetic field, respectively and
\begin{eqnarray}
&&U(\sigma)= \frac{1}{2}m_\sigma^2\sigma^2+\frac{1}{3}g_2\sigma^3+\frac{1}{4}g_3\sigma^4 \ ,\nonumber \\
&&U_\omega(\omega_\mu)= \frac{1}{2}m_\omega^2\omega_\mu\omega^\mu +\frac{1}{4}c_3(\omega_\mu\omega^\mu)^2  \ ,\nonumber \\
&&U_\rho(\vec{\rho}_\mu)= \frac{1}{2}m_\rho^2\vec\rho_\mu\vec\rho^\mu +\frac{1}{4}d_3(\vec\rho_\mu\vec\rho^\mu)^2    \ ,\
\rho=\vec\rho\cdot\vec\tau \ .
\end{eqnarray}
We refer to the table 2 in Ref.~\cite{Meng:2005jv} for the value of the masses and coupling constants in the above Lagrangian 
that were determined by studying the properties of nuclear matter and a few doubly magic nuclei with no-sea and mean-field approximations.
After taking the mean field approximation on the above Lagrangian and performing the Legendre transformation, we obtain the corresponding mean field Hamiltonian $\mathcal{H}_{\rm RMF}$ and the energy density functional 
$E_{\rm RMF}=\langle \Phi| \mathcal{H}_{\rm RMF}|\Phi \rangle$. Here, $|\Phi\rangle$ is the ground state of a nucleus with the mass number $A$, 
$|\Phi\rangle={\displaystyle\prod_{a=1}^A} c_a^\dagger|0\rangle$ and $c_a^\dagger$ is the creation operator of the nucleon field,
$\psi(x)={\displaystyle\sum_a}\psi_a(x)c_a$. 
Then, the expectation value of the Hamiltonian with the mean field approximation reads 
\begin{eqnarray}
E_{\rm RMF} (\rho, \phi) \!\!\!\! & = &\!\!\!\! \langle \Phi| \mathcal{H}_{\rm RMF}|\Phi \rangle\nonumber\\
  \!\!\! & =&\!\!\!\int d^3x{\rm Tr} [ \beta(\vec\gamma\cdot\vec p+ M+g_\sigma\sigma+g_\omega \beta\omega_0 
+g_\rho \beta\rho_0^3\tau_3 +\,\,e\!\!\not\!\! A \frac{1-\tau_3}{2} )  \rho   ] \nonumber \\ 
&&\!\!\!\!\!  +\int d^3x [-\frac{1}{2}\partial^i\sigma\partial_i\sigma + U_\sigma(\sigma)+\frac{1}{4}\Omega^{ij}\Omega_{ij}
-U_\omega (\omega_0) +\frac{1}{4}\vec{R}^{ij}\cdot\vec{R}_{ij} -U_\rho (\rho_0^3) \nonumber\\
&&\,\,\,\,\,\,\,\,\,\,\,\,\,\,\,\,\,\,\,\,-\frac{1}{4}F^{0j}F_{0j} ]\ , \label{ERMF}
\end{eqnarray}
where $\rho$ is the density matrix, $\phi$ represents bosonic fields and $\gamma^\mu=(\beta, \beta\vec\alpha)$.
Here, we assumed that the mean field is time-independent. Also, we have applied the fact that the spatial components of the vector
fields are zero in a system with the time reversal symmetry. 
By performing variations on $E_{\rm RMF}$ with respect to $\rho$ and $\phi$,
we obtain the equations for the nucleon and bosons~\cite{Meng:2005jv}.
\begin{eqnarray}
h_D\psi_i(\vec x)=\epsilon_i\psi_i (\vec x) 
\label{DiracE}
\end{eqnarray}
where the Dirac Hamiltonian $h_D$ is given by
\begin{equation}
h_D = \vec{\alpha}\cdot \vec{p}
+ {\beta}[M + S(\mbr)]+ V(\mbr)
\end{equation}
with the scalar $S(\mbr)$ and vector $V(\mbr)$ potentials given by
\begin{equation}
\begin{aligned}
S(\mbr) & = g_\sigma\sigma(\vec x) \ , \nonumber\\
V(\mbr) & = g_\omega\omega_0(\vec x)+g_\rho\tau_3\rho_0^3(\vec x) +\frac{1}{2}e (1-\tau_3)A_0 (\vec x) \ .
\end{aligned}
\end{equation}
In general, the equations of motion for the nucleon moving in the mean field potentials are solved
by using the the harmonic oscillator basis.
However, for exotic nuclei whose density profile can have a long tail, it is preferable to solve the equations in coordinate space
and adopt a basis  which can treat the asymptotic behavior of the nucleon wave function.  
In Ref.~\cite{Meng:2005jv}, the Woods–Saxon basis was used to solve the equations of motion for the nucleon.

Similarly, by doing variations on $E_{\rm RMF}$ with respect to   $\phi$,
we obtain the equations for the boson~\cite{Meng:2005jv},
\begin{eqnarray}
&&- {\vec\triangledown}^2  \sigma+U_\sigma^\prime (\sigma) =-g_\rho\rho_3\, , \nonumber\\
&&- {\vec\triangledown}^2  +U_\omega^\prime (\omega_0) = g_\omega\rho_\omega \, ,  \nonumber\\
&&- {\vec\triangledown}^2   +U_\rho^\prime(\rho_0^3) = g_\rho\rho_3 \, ,  \nonumber\\
&&- {\vec\triangledown}^2     A_0=e\rho_c\, .  \label{BosonE}
\end{eqnarray}
where
\begin{eqnarray}
&&\rho_s={\rm Tr}[\beta\rho]\, , \nonumber\\
&&\rho_\omega={\rm Tr}[\rho]\, , \nonumber\\
&&\rho_3={\rm Tr}[\tau_3\rho]\, , \nonumber\\
&&\rho_c={\rm Tr}[(1-\tau_3)\rho]\, . 
\end{eqnarray}
Using Eq.~(\ref{BosonE}) in Eq.~(\ref{ERMF}), one can obtain the total energy of the system as
\begin{eqnarray}
E &\!\!\!\!\! =\!\!\!\!\! & \int d^3x{\rm Tr} [ \beta(\vec\gamma\cdot\vec p+ M)\rho+\frac{1}{2}(g_\sigma\beta\sigma+g_\omega \omega_0 
+g_\rho \rho_0^3\tau_3 +A_0\frac{1-\tau_3}{2} )  \rho   ] \nonumber \\ 
&&\!\!\!\!\!\!   + \int d^3x [U_\sigma(\sigma)-U_\omega(\omega_0)-U_\rho(\rho_0^3) -\frac{1}{2}(\sigma U_\sigma^\prime(\sigma)-\omega_0U_\omega^\prime(\omega_0)
-\rho_0^3U_\rho^\prime(\rho_0^3))]\, .
\end{eqnarray}

%%%%%%%%%%%%%%%%%%%%%%%%%%%%%%%%%%%%%%%
\subsection{Finite nuclei and chiral invariant mass}
%%%%%%%%%%%%%%%%%%%%%%%%%%%%%%%%%%%%%%%
In Section~\ref{sec:PDM MF} we have introduced our parity doublet model and fixed the model parameters using the nuclear matter properties.
We observed that the parity doublet model reproduces reasonably the nuclear matter saturation properties with the chiral invariant nucleon mass $m_0$ in the range of $600$-$900$ MeV. Now, to pin down the value of $m_0$,  we study the properties of nuclei using the parity doublet model in the frame work of a self-consistent relativistic mean field theory.

Using the Lagrangian of our parity doublet model in Eq.(\ref{PDM-L1}), 
we obtain the equations of motion (EoM) for the stationary mean fields $\sigma$, $\omega_0$, $\rho_0^3$ and $A_0$
~\cite{mun2023},
\begin{eqnarray}
\Big(-\vec{\nabla}^2 + m_\sigma^2\Big)\langle \sigma(\vec{x}) \rangle &=& - \bar{N}(\vec{x})N(\vec{x}) \left.\frac{\partial\, m_N( \sigma)}{\partial \sigma }\right|_{\sigma=\langle \sigma(\vec{x}) \rangle}~\nonumber\\
&&+\left(-3 f_\pi \lambda+10 f_\pi^3 \lambda_6\right)\langle \sigma(\vec{x}) \rangle^2 \nonumber\\
&&+\left(-\lambda+10 f_\pi^2\lambda_6\right)\langle \sigma(\vec{x}) \rangle^3 \nonumber\\
&&+5 f_\pi\lambda_6\langle \sigma(\vec{x}) \rangle^4+\lambda_6\langle \sigma(\vec{x}) \rangle^5\, ,\label{EoMsig} \\
\Big(-\vec{\nabla}^2 + m_\omega^2 \Big)\langle \omega_0(\vec{x}) \rangle &=& g_{\omega N\!N}N^\dagger(\vec{x}) N(\vec{x})\,, \label{EoMome}\\
%% - 4g_4^4 \langle \omega_0(\vec{x}) \rangle^3\,, \label{EoMome}\\
\Big(-\vec{\nabla}^2 + m_\rho^2 \Big)\langle \rho_0^3(\vec{x}) \rangle &=& g_{\rho N\!N}N^\dagger(\vec{x})\frac{\tau^3}{2} N(\vec{x})\,, \label{EoMrho}\\
-\vec{\nabla}^2 \langle A_0(\vec{x}) \rangle &=& e N^\dagger(\vec{x}) \frac{1-\tau_3}{2} N(\vec{x})\,. \label{EoMphoton}
\end{eqnarray}
Note here that we take the shift $\sigma \to f_\pi + \sigma$ since the scalar  field in the parity doublet model is a chiral partner of
the pion field whose vacuum expectation value in free space is $f_\pi$, while that of the widely used scalar field in nuclear structure studies is zero in free space.
Since we are interested in finite nuclei, we will not consider the EoM for the parity partner of the nucleon, $N^\ast(1535)$, which does not form its Fermi sea near the saturation density. In addition, since our primary goal here is to see if the parity doublet model can explain some basic nuclear properties such as the binding energy with a reasonable value of the chiral invariant mass, we will not consider pairing correlations which are essential for odd-even staggering in nuclear properties. For instance, according to the semi-empirical mass formula, the contribution from the pairing term to the binding energy per nucleon of $^{58}$Ni is only about $0.03$~MeV.

\begin{figure}[h]
\includegraphics[width=0.5\textwidth]{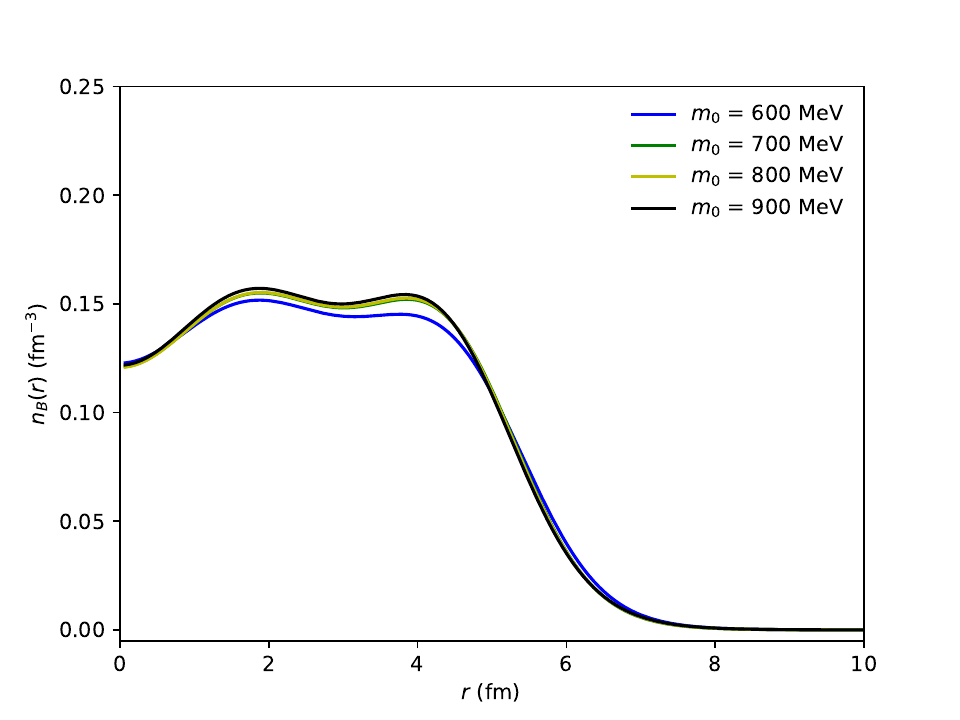} 
\includegraphics[width=0.5\textwidth]{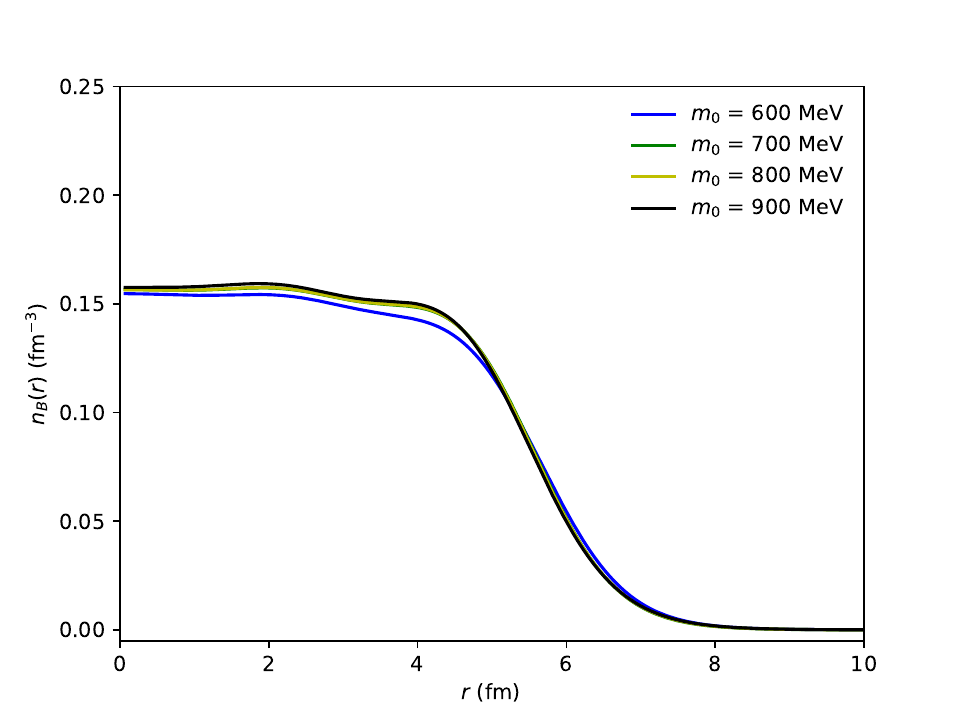}
\caption{%(Color online) 
Nucleon density profile in $^{112}$Sn (on the left) and $^{126}$Sn (on the right) calculated with $K=215$\,MeV.
}\label{fig-density}
\end{figure}

The EoM for the nucleon is given by
\begin{equation}
\big[ \vec{\alpha} \cdot \vec{p} + \beta\, m_N(\langle \sigma(\vec{x}) \rangle)  +V(\vec{x}) \big]N_i(\vec{x}) = \epsilon_i N_i(\vec{x})\ , \label{nucleon}
\end{equation}
where $N_i$ is the single-particle wave function and
\begin{equation}
V(\vec{x})=g_{\omega N\!N} \langle\omega_0(\vec{x})\rangle + g_{\rho N\!N}\langle\rho_0^3(\vec{x})\rangle\frac{\tau^3}{2}+e \frac{(1-\tau_3 )}{2} \langle A_0(\vec{x})\rangle\ .
\end{equation}
With assuming the spherical shape of the nucleus, we can solve Eqs.~(\ref{EoMsig})-(\ref{EoMphoton}) and Eq.~(\ref{nucleon})   simultaneously to obtain the energy
\begin{equation}
E = \int d^3x \, {\cal H}(\vec{x}) \ .
\end{equation}
After subtracting out the vacuum contribution,  we write the Hamiltonian density ${\cal H}(\vec{x})$  in the mean field approximation as
%\begin{widetext}
\begin{eqnarray}
{\cal H} &=& \bar{N} \left( -i\gamma^i \partial_i + m_N \right) N + g_{\omega N\!N} \langle \omega_0 \rangle N^\dagger N + g_{\rho N\!N} \langle \rho_0^3 \rangle N^\dagger \frac{\tau^3}{2} N + e \langle A_0 \rangle N^\dagger \frac{1-\tau_3}{2}N \nonumber \\
&&-\frac{1}{2} \partial^i\langle \sigma \rangle \partial_i \langle \sigma \rangle +\frac{1}{2} \partial^i \langle \omega_0 \rangle \partial_i \langle \omega_0 \rangle +\frac{1}{2} \partial^i \langle \rho_0^3 \rangle \partial_i \langle \rho_0^3 \rangle +\frac{1}{2} \partial^i \langle A_0 \rangle \partial_i \langle A_0 \rangle \nonumber\\
&& -\frac{\bar{\mu}^2}{2} \left[ \left( f_\pi +\langle \sigma \rangle \right)^2 -f_\pi^2 \right] + \frac{\lambda}{4} \left[ \left( f_\pi +\langle \sigma \rangle \right)^4 -f_\pi^4 \right] - \frac{\lambda_6}{6} \left[ \left( f_\pi +\langle \sigma \rangle \right)^6 -f_\pi^6 \right] -\epsilon \langle \sigma \rangle \nonumber\\
&&-\frac{1}{2}m_\omega^2 \langle \omega_0 \rangle^2 -\frac{1}{2}m_\rho^2 \langle \rho_0^3 \rangle^2\,. \label{Hamil}
\end{eqnarray}
%\end{widetext}
Then, the binding energy (BE) per nucleon  is given by
\begin{equation}
{\rm BE}/{A} = -\frac{E}{A} +m_N\,.
\end{equation}
To put an additional constraint on the value of the chiral invariant mass, using the model parameters summarized in Table~\ref{PD1215}, we calculate the binding energies per nucleon and charge radii of selected nuclei: $^{16}$O,  $^{40}$Ca, $^{48}$Ca, $^{58}$Ni, $^{70}$Ge, $^{82}$Se, $^{92}$Mo, $^{112}$Sn, $^{126}$Sn, $^{138}$Ba, $^{154}$Sm, $^{170}$Er, $^{182}W$, $^{202}$Pb and $^{208}$Pb~\cite{mun2023}. 
Before we compare our results for the binding energies and charge radii with the experiments, we show the nucleon density profile,
mean-field value and effective nucleon mass in a nucleus with different values of the chiral invariant mass to visualize how the chiral invariant mass affects them.
We first plot the nucleon density profile in $^{112}$Sn 
and $^{126}$Sn for different values of the chiral invariant mass in Fig.~\ref{fig-density}. It is interesting to see that  $^{112}$Sn has a depleted central density and therefore can be a candidate of bubble nuclei, which was also observed in the previous studies based on relativistic mean field models, for example see Ref.~\cite{Sunardi:2018snn}. 
%
%%%%%%%
%%%%%%%

\begin{figure}[h]
\includegraphics[width=0.5\textwidth]{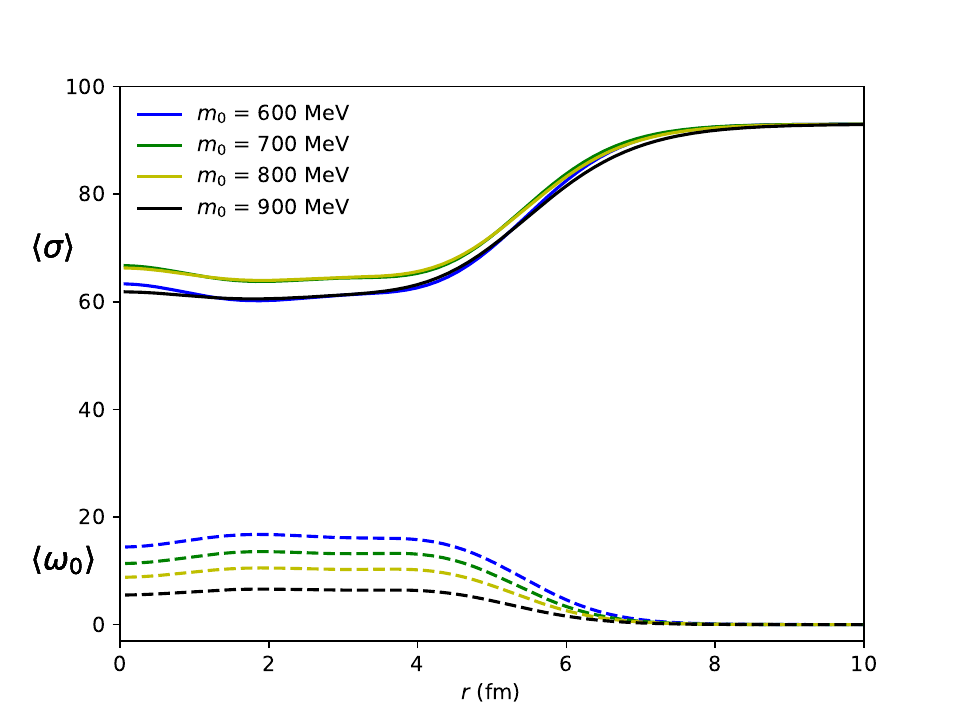}
\includegraphics[width=0.5\textwidth]{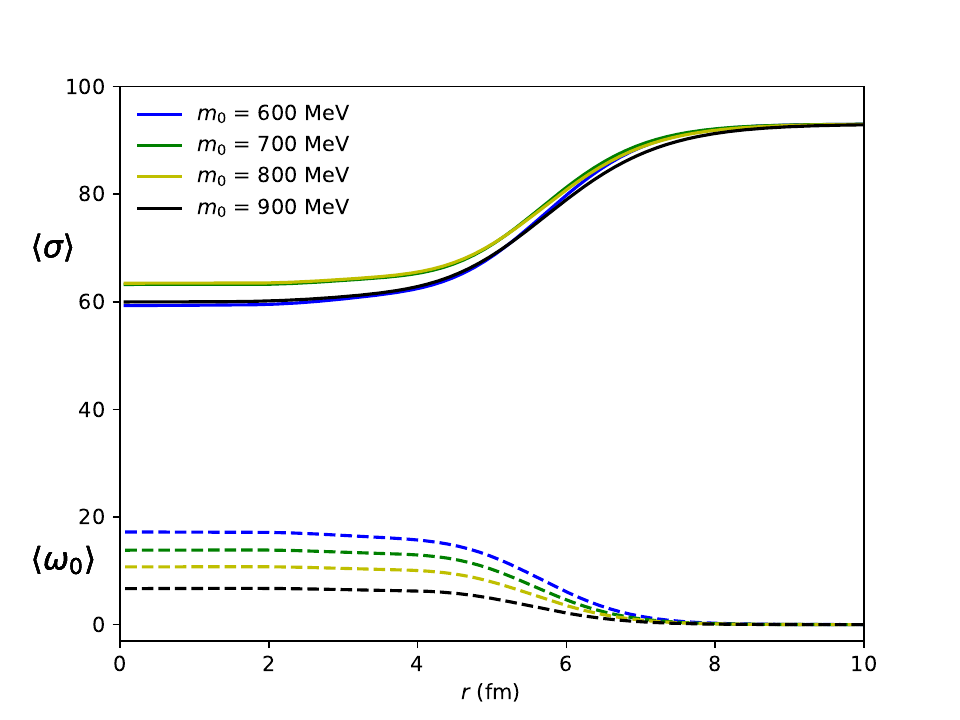}
\caption{%(Color online) 
$\langle\sigma\rangle$ and $\langle\omega_0\rangle$  in $^{112}$Sn (on the left) and $^{126}$Sn (on the right) with $K=215$\,MeV.
}\label{fig-SO}
\end{figure}

In Fig~\ref{fig-SO}, we present the value of $\langle\sigma\rangle$ and $\langle\omega_0\rangle$ in $^{112}$Sn 
and $^{126}$Sn for different values of the chiral invariant mass. As expected, the value of $\langle\sigma\rangle$ decreases and $\langle\omega_0\rangle$ increases as $r\rightarrow 0$, from zero density to the saturation density. 
%
%%%%%%%
%%%%%

\begin{figure}[h]
\centering
\includegraphics[width=0.55\textwidth]{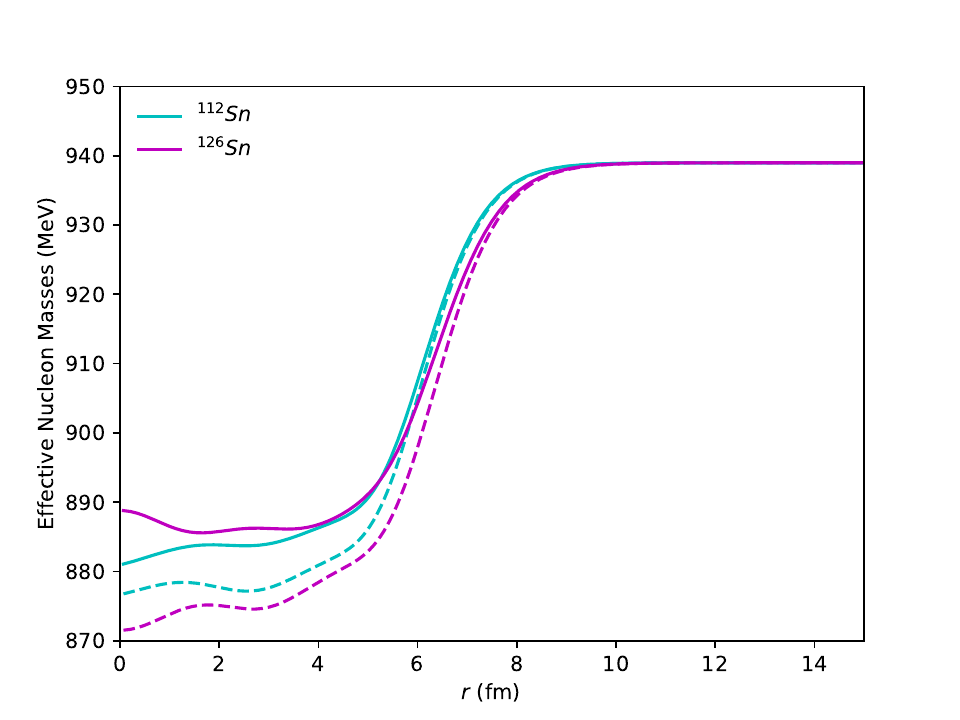}
\caption{%(Color online) The 
Neutron and proton masses in $^{112}$Sn and $^{126}$Sn with $K=215$\,MeV.
Here, solid (dashed) lines are for the neutron (proton).
}\label{fig-NP}
\end{figure}

The effective neutron and proton  masses in $^{112}$Sn and $^{126}$Sn are shown in Fig.~\ref{fig-NP}, where the effective mass is defined as  the energy of the nucleon at rest:
\begin{eqnarray}
&& m_n^{\rm (eff)} = m_n+g_{\omega NN}\langle\omega_0\rangle-\frac{g_{\rho NN}}{2}\langle\rho_0\rangle\, ,\nonumber \\
&& m_p^{\rm (eff)} = m_p+g_{\omega NN}\langle\omega_0\rangle+\frac{g_{\rho NN}}{2}\langle\rho_0\rangle\, . \nonumber
\end{eqnarray}
As observed in Ref.~ \cite{mun2023},  the neutron-proton mass difference becomes larger in a nucleus with larger isospin asymmetry.
%%%%%%%
%%%%%%%

Now, as in Ref.~ \cite{mun2023} we  compare our results with experiments to check which value of the chiral invariant mass reproduces well the experimental results. 
In Table~\ref{RMS-deviation}, we present the root-mean-square deviation (RMSD) of the binding energies and  charge radii for $m_0=600$, $700$, $800$, $900$\,MeV. As concluded in Ref.~\cite{mun2023}, it can be seen from Table~\ref{RMS-deviation} that the case with $m_0=700$\,MeV is preferred by the nuclear properties considered in this work. 
\begin{table}[htbp] 
\caption{Root-mean-square deviations for $m_0=600$, $700$, $800$, $900$\,MeV with two different values of  the incompressibility.}
\label{RMS-deviation}
\newcolumntype{C}{>{\centering\arraybackslash}X}
\begin{tabularx}{\textwidth}{CCCCCC}
\toprule
 &\textbf{$m_0$ (MeV)}	& \textbf{600}	& \textbf{700}  & \textbf{800}  & \textbf{900}\\
\midrule
\multirow[m]{2}{*}{$K_0 = 215$ MeV} & RMSD: BE/A (MeV) & 0.573 &0.438 &0.727 & 1.734 \\
& RMSD: $R_c$ (fm) &~ 0.082 &~0.046  &~0.049 &~ 0.056~  \\
\midrule
\multirow[m]{2}{*}{$K_0 = 240$ MeV} & RMSD: BE/A (MeV) &~ 0.827 &~0.737 &~1.047 &~  2.147~ \\
& RMSD: $R_c$ (fm) &~ 0.097 &~0.052  &~0.053 &~ 0.062~  \\
\bottomrule
\end{tabularx}
\end{table}

%%%%%%%%%%%%%%%%%%%%%%%%%%%%%%%%%%%%%%%%%%
%\section{Infinite nuclear matter}
\section{Effect of $a_0$ meson in  nuclear matter}
\label{sec:a0}

\subsection{Extended parity doublet model}

In the previous sections, the scalar field $M$ includes only $\sigma$ and $\pi$ fields. However, the isovector scalar meson may be important in particular when we study the asymmetric matter such as neutron star matter and neutron-rich nuclei. In Ref.~\cite{PhysRevC.108.055206}, the PDM is extended to include the lightest isovector scalar meson $a_0(980)$ and its chiral partner $\eta$ meson by reparametrising the meson field $M$ as
\begin{equation}
    M = [\sigma + i\vec{\pi} \cdot \vec{\tau}] -   [ \vec{a_0} \cdot \vec{\tau} + i\eta]\ ,
    \label{Ma}
\end{equation}
where $\vec{a_0}$ is the isovector scalar field corresponding to the $a_0(980)$ meson and $\eta$ the isoscalar pseudoscalar field corresponding to the $\eta$ meson.
The potential is written as
\begin{align}
V_M=&-\frac{\bar{\mu}^2}{4}\mbox{tr}[M^\dagger M ]+\frac{\lambda_{41}}{8}\mbox{tr}[(M^\dagger M)^2] \nonumber \\ 
&{}-\frac{\lambda_{42}}{16}\{ \mbox{tr}[M^\dagger M]\}^2-\frac{\lambda_{61}}{12}\mbox{tr}[(M^\dagger M)^3] \nonumber \\ 
&{}-\frac{\lambda_{62}}{24}\mbox{tr}[(M^\dagger M)^2]\mbox{tr}[M^\dagger M]-\frac{\lambda_{63}}{48}\{ \mbox{tr}[M^\dagger M]\}^3 \nonumber \\
&{}-\frac{m^2_{\pi}f_{\pi}}{4}\mbox{tr}[M+M^\dagger]-\frac{K}{8}\{\det M+\det M^\dagger\} \ ,
\label{VM}
\end{align} 
where we included all the terms invariant under SU(2)$_L\times$SU(2)$_R\times$U(1)$_A$ symmetry up to sixth order in $M$ fields.
The term proportional to $m_\pi^2$ is the explicit chiral symmetry breaking term which generates the mass of pion.
The last term is the Kobayashi-Maskawa-’t~Hooft interaction term introduced to account for the U(1)$_A$ anomaly.

We note that the sub-leading order terms in large $N_c$ expansion, i.e., $\lambda_{42}$, $\lambda_{62}$, $\lambda_{63}$ terms as well as the $K$ term are not independent when only the $\sigma$ and $\pi$ fields exist as in section~\ref{sec:PDM MF}, while they are independent in this section due to the inclusion of $a_0$ and $\eta$ fields.

Under the mean field approximation, the $a_0$ meson has a mean field given by
\begin{equation}
    a_0^i(x) \rightarrow a\,\delta_{i3} \ ,  %, \qquad a_0^{i=3} \equiv a\ .
\end{equation} 
with the $\eta$ meson mean field assumed to be zero in dense matter, similar to the pion field:
\begin{equation}
\eta(x) \rightarrow 0\ .
\end{equation} 
Then, the mean field for $M$ becomes
\begin{equation}
\left\langle M \right\rangle = \begin{pmatrix}
\sigma - a & 0 \\ 0 & \sigma + a
\end{pmatrix}
\ .
\end{equation}
As a result, the mean field potential is expressed as
\begin{equation}
\begin{aligned}
V_M = &  - \frac{\bar{\mu}^2_{\sigma}}{2} \sigma^2 - \frac{\bar{\mu}^2_{a}}{2} a^2  +  \frac{\lambda_4}{4}  (\sigma^4 + a^4 )  + \frac{\gamma_4}{2} \sigma^2 a^2 \\ 
& {} - \frac{\lambda_6}{6}  (\sigma^6 +15\sigma^2a^4 + 15\sigma^4a^2 +a^6 )  +\lambda_6^{'}(\sigma^2a^4 + \sigma^4a^2) \\ 
& {}  - m^2_{\pi} f_{\pi} \sigma \ ,
\end{aligned}
\label{VMa}
\end{equation} 
where the parameters are defined as 
\begin{equation}
\begin{aligned}
{} & \bar{\mu}_{\sigma}^2 \equiv \bar{\mu}^2 + \frac{1}{2} K\ , \\
& \bar{\mu}^2_a \equiv \bar{\mu}^2 - \frac{1}{2} K  = \bar{\mu}_{\sigma}^2 - K\ , \\
& \lambda_4 \equiv \lambda_{41} - \lambda_{42}\ , \\
& \gamma_4 \equiv 3\lambda_{41} - \lambda_{42}\ , \\
& \lambda_6  \equiv \lambda_{61} + \lambda_{62} + \lambda_{63}\ , \\
& \lambda_6^{'}  \equiv \frac{4}{3}\lambda_{62} + 2\lambda_{63}\ . 
\label{eq2.23}
\end{aligned}
\end{equation}
In the present model, $\lambda_6^{'}$ is taken as a free parameter to examine the effect of $\lambda_{62}$ and $\lambda_{63}$ interactions, which are of sub-leading order in the large $N_c$ expansion. Given that the $ \lambda'_6 $ term is suppressed by $ 1/N_c $ compared to the $ \lambda_6 $ term in the large $ N_c $ expansion, we assume $ |\lambda'_6| \lesssim |\lambda_6| $ holds. Consequently, we consider $ \lambda'_6 = 0, \pm \lambda_6 $ to assess the impact of the sub-leading order six-point interactions on the symmetry energy. By default, we first set $ \lambda'_6 = 0 $. In the end of sub-sections~\ref{Sa0woVM} and \ref{Sa0wVM}, we investigate the impact of $\lambda'_6$ on the results by comparing the cases with $ \lambda'_6 = 0, \pm \lambda_6 $.

In the present model, the vector mesons $\rho^\mu$ and $\omega^\mu$ are included based on the HLS with polar decomposition of the field $M$ as 
\begin{equation}
\begin{aligned}
M = \xi_L^\dagger S \xi_R \, ,
\end{aligned}
\label{HLSM}
\end{equation}
where 
$S = \sigma + \sum_{b=1}^3 {a}_0^b \tau_b/2$ is the $2\times2$ matrix field for scalar mesons.
$\xi_{L,R}$ are transformed as 
\begin{equation}
\begin{aligned}
\xi_{L,R} \rightarrow h_\omega h_\rho \xi_{L,R} g_{L,R}^\dagger \ ,
\end{aligned}
\label{HLSXItr}
\end{equation}
with $h_\omega \in U(1)_{HLS}$ and $h_\rho \in SU(2)_{HLS}$. In the unitary gauge, $\xi_{L,R}$ are parameterized as
\begin{equation}
%\begin{aligned}
\xi_R = \xi_L^\dagger %= exp(i P_a T^a/f_\pi),
= \exp \left( i P / f_\pi\right) \ ,
%\end{aligned}
\label{HLSXI}
\end{equation}
where 
$P = i\eta + \sum_{a=1}^{3} \pi^a \tau_a/2$
is the $2\times2$ matrix field for 
pseudoscalar mesons.
In the HLS, vector mesons are introduced as the gauge bosons of the HLS. The vector mesons then transform as  
\begin{equation}
\begin{aligned}
\omega_\mu \rightarrow h_\omega \omega_\mu h_\omega^\dagger + \frac{i}{g_\omega}\partial_\mu h_\omega h_\omega^\dagger \ ,
\end{aligned}
\label{HLSomega}
\end{equation}
\begin{equation}
\begin{aligned}
\rho_\mu \rightarrow h_\rho \rho_\mu h_\rho^\dagger + \frac{i}{g_\rho}\partial_\mu h_\rho h_\rho^\dagger \ ,
\end{aligned}
\label{HLSrho}
\end{equation}
where $g_\omega$ and $g_\rho$ are the corresponding gauge coupling constants. 
The
covariantized Maurer-Cartan 1-forms
are given by 
\begin{equation}
    \hat{\alpha}_\perp^\mu \equiv \frac{1}{2i}[D^\mu \xi_R \xi_R^\dagger - D^\mu \xi_L \xi_L^\dagger] \ ,
\end{equation}
\begin{equation}
    \hat{\alpha}_\parallel^\mu \equiv \frac{1}{2i}[D^\mu \xi_R \xi_R^\dagger + D^\mu \xi_L \xi_L^\dagger] \ ,
\end{equation}
 where
the covariant derivatives
are defined as 
\begin{align}
   D^\mu \xi_L =  \partial^\mu \xi_L - ig_\rho \rho^\mu \xi_L - ig_\omega \omega^\mu \xi_L  + i \xi_L \mathcal{L}^\mu - i \xi_L \mathcal{A}^\mu \ ,
\\
   D^\mu \xi_R =  \partial^\mu \xi_R - ig_\rho \rho^\mu \xi_R - ig_\omega \omega^\mu \xi_R + i \xi_R \mathcal{R}^\mu + i \xi_L \mathcal{A}^\mu \ ,
\end{align}
with $\mathcal{L}^\mu$, $\mathcal{R}^\mu$ and $\mathcal{A}^\mu$ being the external gauge fields corresponding to SU(2)$_L \times$SU(2)$_R \times$U(1)$_A$ global symmetry.
\footnote{
We note that mesons do not carry the baryon number, so that the external gauge field corresponding to U(1) baryon number does not appear in the above covariant derivative.
We also note that the covariant derivative acting on the baryon fields includes the external gauge field $\mathcal{A}^\mu$. Anyway, $\mathcal{A}^\mu$ has no contribution in dense matter in the mean field approximation.
}
Now, 
the HLS-invariant Lagrangian is given by 
\begin{equation}
\begin{aligned}
\mathcal{L}_{V} = {} & a_{VNN} \left[ \Bar{N}_{1l} \gamma^\mu \xi_L^\dagger \hat{\alpha}_{\parallel \mu} \xi_L N_{1l} + \Bar{N}_{1r} \gamma^\mu \xi_R^\dagger \hat{\alpha}_{\parallel \mu} \xi_R N_{1r} \right] \\ 
    & + a_{VNN} \left[ \Bar{N}_{2l} \gamma^\mu \xi_R^\dagger \hat{\alpha}_{\parallel \mu} \xi_R N_{2l} + \Bar{N}_{2r} \gamma^\mu \xi_L^\dagger \hat{\alpha}_{\parallel \mu} \xi_L N_{2r} \right] \\ 
    &  + a_{0NN} \sum_{i = 1,2} \left[ \Bar{N}_{il} \gamma^\mu \mbox{tr} [ \hat{\alpha}_{\parallel \mu} ] N_{il} + \Bar{N}_{ir} \gamma^\mu  \mbox{tr} [\hat{\alpha}_{\parallel \mu} ] N_{ir} \right] \\
& + \frac{{m_\rho}^2}{{g_\rho}^2}\mbox{tr} [\hat{\alpha}_\parallel^\mu \hat{\alpha}_{\parallel \mu}]  +  \left(\frac{{m_\omega}^2}{8{g_\omega}^2} - \frac{{m_\rho}^2}{2{g_\rho}^2}\right) \mbox{tr} [\hat{\alpha}_\parallel^\mu]\mbox{tr}[ \hat{\alpha}_{\parallel \mu}] - \frac{1}{8{g_\omega}^2}\mbox{tr}[\omega^{{\mu\nu}}\omega_{{\mu\nu}}] - \frac{1}{2{g_\rho}^2}\mbox{tr}[\rho^{{\mu\nu}}\rho_{{\mu\nu}}] \\
& + \lambda_{\omega \rho}\left( a_{VNN} + a_{0NN} \right)^2 a_{VNN}^2 \left[ \frac{1}{2}\mbox{tr} [\hat{\alpha}_\parallel^\mu \hat{\alpha}_{\parallel \mu}]\mbox{tr} [\hat{\alpha}_\parallel^\nu]\mbox{tr}[ \hat{\alpha}_{\parallel \nu}] - \frac{1}{4}\left\{ \mbox{tr} [\hat{\alpha}_\parallel^\mu]\mbox{tr}[ \hat{\alpha}_{\parallel \mu}] \right\}^2 \right] .
\
\end{aligned}
\label{Lv}
\end{equation}
In particular, the last term is a mixing term of $\rho$ and $\omega$ mesons as introduced in Ref.~\cite{PhysRevC.108.055206} to the $a_0$ model to reduce the slope parameter, following Ref.~\cite{PhysRevC.106.065205}. As we will show in sub-section~\ref{Sa0woVM}, when we just add the effect of $a_0(980)$ meson to the PDM without this vector meson mixing term, the slope parameter becomes too large compared with the recent constraints as summarized in Ref.~\cite{universe7060182}.

Expressing vector mesons
in terms of mean fields, the Lagrangian becomes
\begin{equation}
\begin{aligned}
\mathcal{L}_{V} = & - g_{\omega NN} \sum_{\alpha j} \Bar{N}_{\alpha j} \gamma^0 \omega N_{\alpha j} - g_{\rho NN} \sum_{\alpha j} \Bar{N}_{\alpha j} \gamma^0  \frac{\tau_3}{2} \rho N_{\alpha j} \\
& + \frac{1}{2} m^2_{\omega} \omega^2 + \frac{1}{2} m^2_{\rho} \rho^2 + \lambda_{\omega \rho} g_{\omega NN}^2 g_{\rho NN}^2 \omega^2 \rho^2\ . 
\end{aligned}
\end{equation} 
with
\begin{align}
    g_{\omega NN} & =  \left(  a_{VNN} + a_{0NN}  \right) g_\omega \ ,\notag \\
    g_{\rho NN} & =  a_{VNN} g_\rho \ .
\end{align}

It is crucial to note that $\lambda_{\omega \rho} > 0$ is required to realize $\omega = \rho = 0$ in vacuum. To show it, we start from the vector meson potential in vacuum given as
\begin{equation}
\begin{aligned}
V_{V} \equiv  -\frac{1}{2} m^2_{\omega} \omega^2 - \frac{1}{2} m^2_{\rho} \rho^2 - \lambda_{\omega \rho} g_{\omega NN}^2 g_{\rho NN}^2 \omega^2 \rho^2 \ .
\end{aligned}
\end{equation} 
The vacuum expectation values of the vector meson fields are chosen at the stationary point of $V_{V}$ with minimal energy.
 The stationary conditions are given by:
\begin{equation}
\begin{aligned}
\frac{\partial V_{V}}{\partial \omega} = \omega[m^2_{\omega}  + 2\lambda_{\omega \rho} g_{\omega NN}^2 g_{\rho NN}^2  \rho^2 ] = 0 \ , \\ 
\frac{\partial V_{V}}{\partial \rho} = \rho[m^2_{\rho}  + 2\lambda_{\omega \rho} g_{\omega NN}^2 g_{\rho NN}^2  \omega^2 ] = 0 \ ,
\end{aligned}
\end{equation} 
leading to two distinct stationary points:
\begin{equation}
\begin{aligned}
(\omega^2,\rho^2) = (0,0) , (-\frac{m^2_{\rho}}{2\lambda_{\omega \rho} g_{\omega NN}^2 g_{\rho NN}^2},-\frac{m^2_{\omega}}{2\lambda_{\omega \rho} g_{\omega NN}^2 g_{\rho NN}^2})\ .
\end{aligned}
\end{equation} 
Then, the values of potential at stationary points are
\begin{equation}
V_{V}=
\begin{cases}
      0, & \text{for}\ (\omega^2,\rho^2) = (0,0) \ , \\
      \frac{m^2_{\omega}m^2_{\rho}}{4\lambda_{\omega \rho} g_{\omega NN}^2 g_{\rho NN}^2}, & \text{for}\ (\omega^2,\rho^2) = (-\frac{m^2_{\rho}}{2\lambda_{\omega \rho} g_{\omega NN}^2 g_{\rho NN}^2},-\frac{m^2_{\omega}}{2\lambda_{\omega \rho} g_{\omega NN}^2 g_{\rho NN}^2})\ .
    \end{cases}
  \end{equation} 
In the present model, vanishing vacuum expectation values of the vector meson fields are required at zero density due to the Lorentz-invariance of the vacuum. Consequently, we must require $\lambda_{\omega \rho} > 0$ here such that $(\omega^2,\rho^2) = (0,0)$ minimizes the effective potential $V_{V}$ in vacuum.

The new thermodynamic potential is now given by 
\begin{equation}
\begin{aligned}
{} & \Omega_H   = \Omega_{N} \\
& \qquad \;\;  - \frac{\bar{\mu}^2_{\sigma}}{2} \sigma^2 - \frac{\bar{\mu}^2_{a}}{2} a^2  +  \frac{\lambda_4}{4}  (\sigma^4 + a^4 ) + \frac{\gamma_4}{2} \sigma^2 a^2  \\
& \qquad \;\; - \frac{\lambda_6}{6}  (\sigma^6 +15\sigma^2a^4 + 15\sigma^4a^2+ a^6 )   + \lambda_6^{'}(\sigma^2a^4 + \sigma^4a^2) \\ 
& \qquad \;\;  - m^2_{\pi} f_{\pi} \sigma - \frac{1}{2} m^2_{\omega} \omega^2 - \frac{1}{2} m^2_{\rho} \rho^2 - \lambda_{\omega \rho} g_{\omega NN}^2 g_{\rho NN}^2 \omega^2 \rho^2\\
& \qquad \;\; - \Omega_{0}\ ,
\end{aligned}
\label{eq36}
\end{equation} 
with the new mean field values and the effective nucleon mass given by
\begin{equation}
\begin{aligned}
    m^*_{\alpha j} =  \frac{1}{2} \bigg[ \sqrt{(g_1+g_2)^2(\sigma - ja)^2 + 4m_0^2} + \alpha(g_1 - g_2)(\sigma - ja) \bigg]\ ,
\end{aligned}
\label{maj}
\end{equation} 
where $\alpha=\pm$ referes to the parity, and $j=\pm$ to the iso-spin ($j=+$ for proton and $j=-$ for neutron). We note that the masses of proton and neutron become non-degenerate in the asymmetric matter due to the non-zero mean field of $a_0(980)$.

\subsection{Determination of the model parameters}

In the present model, there are seven parameters in the meson potential,
$\mu_\sigma^2$, $\mu_a^2 = \mu_\sigma^2 - K$, $\lambda_4$, $\gamma_4$, $\lambda_6$, $\lambda'_6$, and $\lambda_{\omega\rho}$, in addition to the meson masses $m_\pi$, $m_\omega$, $m_\rho$ and the pion decay constant $f_\pi$. We also have four parameters, $g_1$, $g_2$, $g_{\omega NN}$ and $g_{\rho NN}$ for the couplings of mesons to baryons. As in section~\ref{sec:PDM MF}, we use the physical values of three masses $m_\pi$, $m_\omega$ and $m_\rho$ together with the pion decay constant $f_\pi$ as listed in Table~\ref{vacM}. Similar to Section~\ref{sec:PDM MF}, we determine the values of $\mu_\sigma^2$, $\lambda_4$, $\lambda_6$ and $g_{\omega NN}$ from the saturation properties: the saturation density $n_0$, the binding energy $B_0$ and the incompressibility $K_0$ 
summarized in Table~\ref{SP},
combined with the vacuum condition given in Eq.~(\ref{vacuum condition}). $g_1$ and $g_2$ is determined from the vacuum mass of nucleon $N(939)$ and its parity partner $N^*(1535)$. The resultant values are same as those shown in 
Table~\ref{PD1215}. 
Then, the parameters $K$ and $\gamma_4$ are determined from the meson masses and the other parameters as
\begin{align}
K = & \quad m_\eta^2 - m_\pi^2 \ , \notag\\
\gamma_4 = & \quad \frac{m_{a_0}^2 + (5 \lambda_{6} - 2\lambda_{6}') f_{\pi}^4 + \bar{\mu}^2_{a}}{f_{\pi}^2}  \ ,
\end{align}
where $m_\eta$ and $m_{a_0}$ are the masses of $\eta$ and $a_0$, the experimental values of which are listed in Table~\ref{vacMa}.
\begin{table}[htbp] 
\caption{Values of meson masses $m_{a_0}$ and $m_\eta$ in unit of MeV.\label{vacMa}}
\newcolumntype{C}{>{\centering\arraybackslash}X}
\begin{tabularx}{\textwidth}{CC}
\toprule
\textbf{$m_{a_0}$}&
\textbf{$m_\eta$}\\
\midrule
980 & 550\\
\bottomrule
\end{tabularx}
\end{table}

As we stated in the previous subsection, we take $\lambda'_6=0$ for a while. The resultant values of $\bar{\mu}^2_{a}$ and $\gamma_4$ corresponding to a given $m_0$ are presented in Tables~\ref{PD10-215}.
\begin{table}[htbp] 
\caption{Values of paramters $\bar{\mu}_a^2$ and $\gamma_4$ for several choices of $m_0$ and $K_0$ with $\lambda'_6=0$.}
\label{PD10-215}
\newcolumntype{C}{>{\centering\arraybackslash}X}
\begin{tabularx}{\textwidth}{CCCCCC}
\toprule
 &\textbf{$m_0$ (MeV)}	& \textbf{600}	& \textbf{700}  & \textbf{800}  & \textbf{900}\\
\midrule
\multirow[m]{2}{*}{$K_0 = 215$ MeV} & $\bar{\mu}^2_a/f^2_{\pi}$&$-$9.40 & $-$11.79 & $-$19.43 & $-$30.27\\ 
& $\gamma_4$&185.59 & 177.94 & 144.51 & 90.62\\
\midrule
\multirow[m]{2}{*}{$K_0 = 240$ MeV} & $\bar{\mu}^2_a/f^2_{\pi}$&$-$10.95 & $-$13.93 & $-$21.08 & $-$31.24\\ 
& $\gamma_4$&176.81 & 164.81 & 133.38 & 83.05\\
\bottomrule
\end{tabularx}
\end{table}

To demonstrate the effect of $a_0$(980) to the matter, we consider both the $a_0$ model with vector-meson mixing ($\lambda_{\omega\rho} \neq 0$) and that without the mixing 
($\lambda_{\omega\rho} = 0$). In the $a_0$ model without vector meson mixing, parameter $g_{\rho NN}$ is determined from the symmetry energy at the saturation density $S_0$, while the slope parameter $L_0$ is computed from the model as an output. 
\begin{table}[htbp] 
\caption{Values of $g_{\rho NN}$ and slope paramter $L_0$ in the $a_0$ model without vector meson mixing, for several choices of $m_0$ and $K_0$ with $\lambda_6^{'}=0$.}
%\label{tablegra0}
\label{tab:gr 215}
\newcolumntype{C}{>{\centering\arraybackslash}X}
\begin{tabularx}{\textwidth}{CCCCCC}
\toprule
 & \textbf{$m_0$ (MeV)}	& \textbf{600}	& \textbf{700}  & \textbf{800}  & \textbf{900}\\
\midrule
\multirow[m]{2}{*}{$K_0 = 215$\,MeV} & $g_{\rho NN}$ &12.52 & 11.20 & 9.94 & 8.90\\
& $L_0$ (MeV)& 120.14 &105.21 & 97.05 &87.65\\
\midrule
\multirow[m]{2}{*}{$K_0 = 240$\,MeV} & $g_{\rho NN}$ &12.47 & 11.16 & 9.90 & 8.86\\
& $L_0$ (MeV)&126.58  & 108.78 & 98.67 & 87.75\\
\bottomrule
\end{tabularx}
\end{table}
Tables~\ref{tab:gr 215} shows the values of $g_{\rho NN}$ togather with $L_0$. We note that the slope parameters are much larger than the recently accepted value $L_0 = 57.7\pm 19$ MeV. 
For making the slope parameter consistent with this value, we include the vector meson mixing interaction which allows us to effectively reduce $L_0$. In the $a_0$ model with vector meson mixing, the parameters $g_{\rho NN}$ and $\lambda_{\omega\rho}$ are determined by fitting them to the symmetry energy $S_0$ as well as the slope parameter $L_0$. To reproduce the matter for recent accepted value of $L_0 = 57.7\pm 19$ MeV, we compute our results for $L_0 = 40$-$80$\,MeV. 
The resultant parameters are shown in Tables~\ref{tab:gr lor 215} and \ref{tab:lwr lor 215}. Here, we only show the results for $\lambda_6^{'}=0$ because the values of the parameters for $\lambda_6^\prime=\pm \lambda_6$ are similar to the values listed in Tables~\ref{tab:gr lor 215} and \ref{tab:lwr lor 215}.
\begin{table}[H] 
\caption{Values of $g_{\rho NN}$ for several choices of $m_0$, $L_0$, with $\lambda_6'$ = 0.}
%\label{tablegra0}
\label{tab:gr lor 215}
\newcolumntype{C}{>{\centering\arraybackslash}X}
\begin{tabularx}{\textwidth}{CCCCCC}
\toprule
 & \textbf{$m_0$ (MeV)}	& \textbf{600}	& \textbf{700}  & \textbf{800}  & \textbf{900}\\
\midrule
\multirow[m]{5}{*}{$K_0 = 215$ MeV} & $L_0=40$ MeV& 15.34 & 13.78 & 12.59 & 11.42 \\
& $L_0=50$ MeV&14.88 & 13.27 & 11.97 & 10.72 \\
& $L_0=60$ MeV&14.46 & 12.81 & 11.44 & 10.13\\
& $L_0=70$ MeV&14.08 & 12.39 & 10.97 & 9.63 \\
& $L_0=80$ MeV&13.72 & 12.02 & 10.55 & 9.19\\
\midrule
\multirow[m]{5}{*}{$K_0 = 240$ MeV} & $L_0=40$ MeV&15.63 & 13.96 & 12.68 & 11.41\\
& $L_0=50$ MeV&15.14 & 13.42 & 12.04 & 10.7\\
& $L_0=60$ MeV&14.69 & 12.94 & 11.49 & 10.11\\
& $L_0=70$ MeV&14.28 & 12.51 & 11.0 & 9.6\\
& $L_0=80$ MeV&13.9 & 12.11 & 10.58 & 9.16 \\
\bottomrule
\end{tabularx}
\end{table}
\begin{table}[H] 
\caption{Values of $\lambda_{\omega\rho}$ for several choices of $m_0$, $L_0$, with $\lambda_6'$ = 0.}
%\label{tablegra0}
\label{tab:lwr lor 215}
\newcolumntype{C}{>{\centering\arraybackslash}X}
\begin{tabularx}{\textwidth}{CCCCCC}
\toprule
 & \textbf{$m_0$ (MeV)}	& \textbf{600}	& \textbf{700}  & \textbf{800}  & \textbf{900}\\
\midrule
\multirow[m]{5}{*}{$K_0 = 215$ MeV} & $L_0=40$ MeV&0.0254 & 0.0818 & 0.3191 & 2.8164 \\
& $L_0=50$ MeV&0.0222 & 0.0693 & 0.2632 & 2.2253\\
& $L_0=60$ MeV&0.0191 & 0.0567 & 0.2072 & 1.6342 \\
& $L_0=70$ MeV&0.0159 & 0.0442 & 0.1513 & 1.0431 \\
& $L_0=80$ MeV&0.0127 & 0.0316 & 0.0954 & 0.4519\\
\midrule
\multirow[m]{5}{*}{$K_0 = 240$ MeV} & $L_0=40$ MeV&0.0252 & 0.0761 & 0.2914 & 2.4593\\
& $L_0=50$ MeV&0.0223 & 0.065 & 0.2418 & 1.9443 \\
& $L_0=60$ MeV&0.0194 & 0.054 & 0.1921 & 1.4293\\
& $L_0=70$ MeV&0.0165 & 0.0429 & 0.1424 & 0.9142\\
& $L_0=80$ MeV&0.0135 & 0.0318 & 0.0927 & 0.3992\\
\bottomrule
\end{tabularx}
\end{table}

\subsection{Symmetry energy of $a_0$ model without vector meson mixing}
\label{Sa0woVM}

The $a_0$(980) meson should affect the properties of the matter via the asymmetry of the matter. Therefore, we expect that symmetry energy is essential to study 
the effect of $a_0$(980) meson to the matter. In the following, we study  how the inclusion of the $a_0(980)$ meson affects the symmetry energy.

In the present model, the symmetry energy S($n_B$) for a given density $n_B$ is expressed as 
\begin{equation}
\begin{aligned}
    S(n_B) & {} = \frac{n_B}{8} 
    \frac{\partial \mu_I}{\partial n_I} \Bigr|_{\substack{n_I=0}}\\
    & =  \frac{(k^*_+)^2}{6 \mu^*_+}   + \frac{n_B}{2}\frac{(g_{\rho NN}/2)^2}{m_{\rho}^2 } -  \frac{n_B}{4} \frac{m^*_+}{\mu^*_+}\frac{\partial m^*_{+n}}{\partial  n_I}\Bigr|_{\substack{n_I=0}} \ ,
    \label{SrhoBawithoutVM}
\end{aligned}
\end{equation} 
where $\mu^*_+ \equiv \mu^*_p \big|_{\substack{n_I=0}}=\mu^*_n \big|_{\substack{n_I=0}}$ is the effective chemical potential for $N(939)$ in the symmetric matter, $k^*_{+} \equiv \sqrt{(\mu^*_p)^2 - (m^*_{+p})^2}  \big|_{\substack{n_I=0}} = \sqrt{(\mu^*_n)^2 - (m^*_{+n})^2}  \big|_{\substack{n_I=0}}$ the corresponding Fermi momentum, $m^*_{+} \equiv m^*_{+p} \big|_{\substack{n_I=0}} = m^*_{+n} \big|_{\substack{n_I=0}}$ the mass.
In Eq.~(\ref{SrhoBawithoutVM}), there are three conributions to the symmetry energy:
the nucleon contribution, the $\rho$ meson contribution, and the $a_0$ meson contribution. Similar to the model in section~\ref{sec:PDM MF}, the nucleon contribution $S_N(n_B)$ is given by 
\begin{equation}
\begin{aligned}
    S_N(n_B) \equiv  \frac{(k^*_+)^2}{6 \mu^*_+}.
    \label{SrhoN}
\end{aligned}
\end{equation}
We note that, since $S_N(n_B)$ arises from the effective kinetic contribution of a nucleon, $S_N(n_B)$ is not affected by the inclusion of $a_0$ meson and thus is the same as in Figure~\ref{SbN without VM}.

\begin{figure}[htbp]
\centering
\includegraphics[width=10.5 cm]{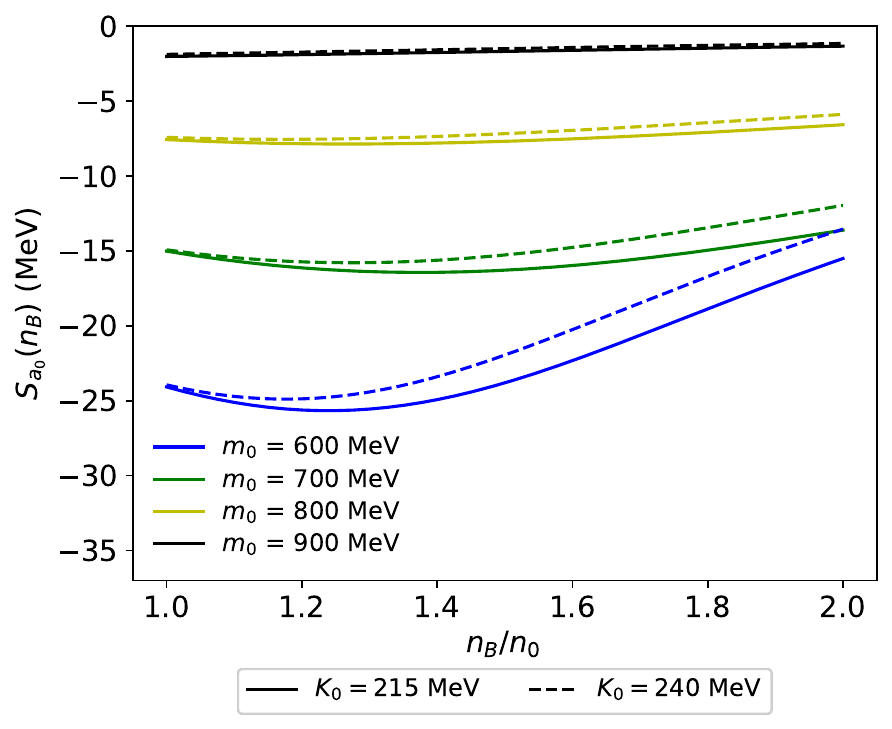}
\caption{$a_0$ meson contribution to the symmetry energy $S_{a_0}(n_B)$ for $m_0 = 600$-$900$\,MeV. Solid curves represent $S_{a_0}(n_B)$ with $K_0 = 215$\,MeV, while dashed curves represent $S_{a_0}(n_B)$ with $K_0 = 240$\,MeV.\label{Sa without VM}}
\end{figure}

\begin{figure}[htbp]
\centering
\includegraphics[width=10.5 cm]{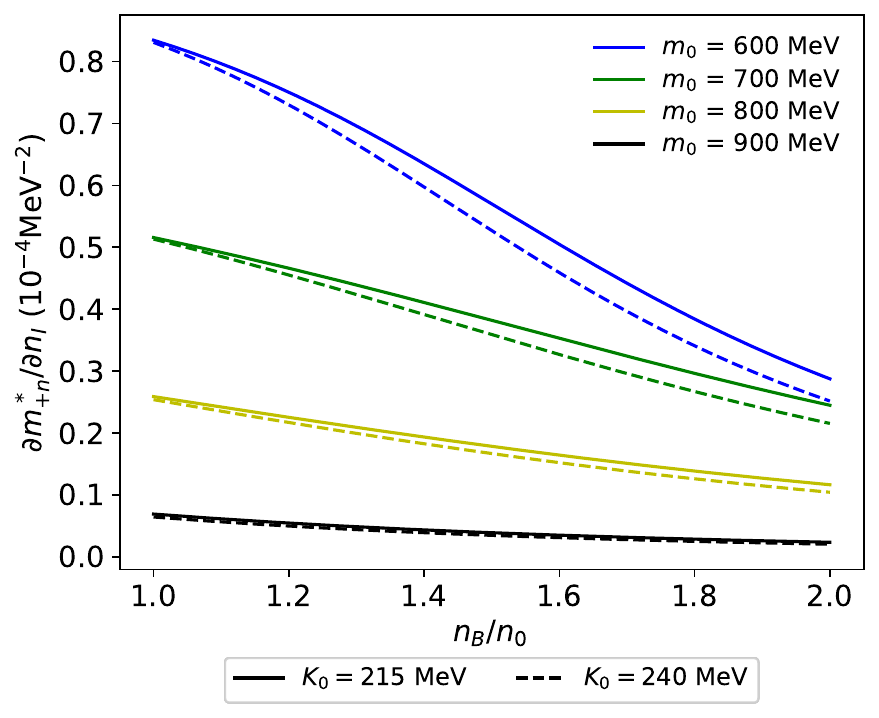}
\caption{$\frac{\partial m^*_{+n}}{\partial n_I}|_{\protect\substack{n_I=0}}$ for $m_0 = 600$-$900$\,MeV. Solid curves represent $\frac{\partial m^*_{+n}}{\partial n_I}|_{\protect\substack{n_I=0}}$ with $K_0 = 215$\,MeV, while dashed curves represent $\frac{\partial m^*_{+n}}{\partial n_I}|_{\protect\substack{n_I=0}}$ with $K_0 = 240$\,MeV.\label{dmdrhoi without VM}}
\end{figure}

The contribution from the $a_0(980)$ meson is expressed as \begin{equation}
\begin{aligned}
    S_{a_0}(n_B)   & {} \equiv  - \frac{n_B}{4} \frac{m^*_+}{\mu^*_+}\frac{\partial m^*_{+n}}{\partial  n_I}\Bigr|_{\substack{n_I=0}}.
    \label{F}
\end{aligned}
\end{equation} 
Figure~\ref{Sa without VM} shows  $S_{a_0}$ computed in the present model.
% %
% \begin{figure}[htbp]
% \centering
% \includegraphics[width=10.5 cm]{Symmetry energy results/Sa without VM.pdf}
% \caption{$a_0$ meson contribution to the symmetry energy $S_{a_0}(n_B)$ for $m_0 = 600$-$900$\,MeV. Solid curves represent $S_{a_0}(n_B)$ with $K_0 = 215$\,MeV, while dashed curves represent $S_{a_0}(n_B)$ with $K_0 = 240$\,MeV.\label{Sa without VM}}
% \end{figure}   
% %
We note that $S_{a_0}$ is negative and thus reduces the total symmetry energy $S(n_B)$. This is because $\frac{\partial m^*_{+n}}{\partial n_I}|_{\substack{n_I=0}}$ is always positive as shown in Figure~\ref{dmdrhoi without VM}.
% %
% \begin{figure}[htbp]
% \centering
% \includegraphics[width=10.5 cm]{Symmetry energy results/dmdrhoi without VM.pdf}
% \caption{$\frac{\partial m^*_{+n}}{\partial n_I}|_{\substack{n_I=0}}$ for $m_0 = 600$-$900$\,MeV. Solid curves represent $\frac{\partial m^*_{+n}}{\partial n_I}|_{\substack{n_I=0}}$ with $K_0 = 215$\,MeV, while dashed curves represent $\frac{\partial m^*_{+n}}{\partial n_I}|_{\substack{n_I=0}}$ with $K_0 = 240$\,MeV.\label{dmdrhoi without VM}}
% \end{figure}   
% %
Intuitively, this can be understood from the dependence of $m^*_{+n}$ on the mean field $a$ given in Eq.~(\ref{maj}).
If we vary the mean field $a$, $m^*_{+n}$ will also change correspondingly. However, the effective chemical potential $\mu_{n}^*$ does not depend on the mean field $a$ directly as we can see from Eq.~(\ref{mustar}). This change of the effective mass $m^*_{+n}$ due to the mean field $a$ leads to the change of the momentum of the neutron $k_{+n} = \sqrt{(\mu_{n}^*)^2 - (m^*_{+n})^2}$. When $n_I = (n_p - n_n)/2$ is increased for a fixed $n_B$, the density of the neutron $n_n$ and thus the momentum $k_{+n}$ is decreased. 
Accordingly, the effective mass of the neutron is increasing as $n_I$ increase, causing a positive $\frac{\partial m_{+n}}{\partial n_I}|_{\substack{n_I=0}}$. Therefore, the $a_0$(980) meson contribution $S_{a_0}(n_B)$ reduces the total symmetry energy $S(n_B)$ in the present model. We also find that the $a_0$(980) effect on the symmetry energy is stronger for smaller $m_0$. This is because the coupling constants of $a_0(980)$ meson to the nucleon, $g_1$ and $g_2$, are larger for smaller $m_0$ as shown in Table~\ref{PD1215}. 
As a result, the symmetry energy becomes larger by $a_0(980)$ meson more when $m_0$ is smaller. In addition, we note that the $a_0(980)$ effect on the symmetry energy is decreasing as the density increases since $\frac{\partial m_{+n}}{\partial n_I}|_{\substack{n_I=0}}$ decreases.

\begin{figure}[h]
\centering
\includegraphics[width=10.5 cm]{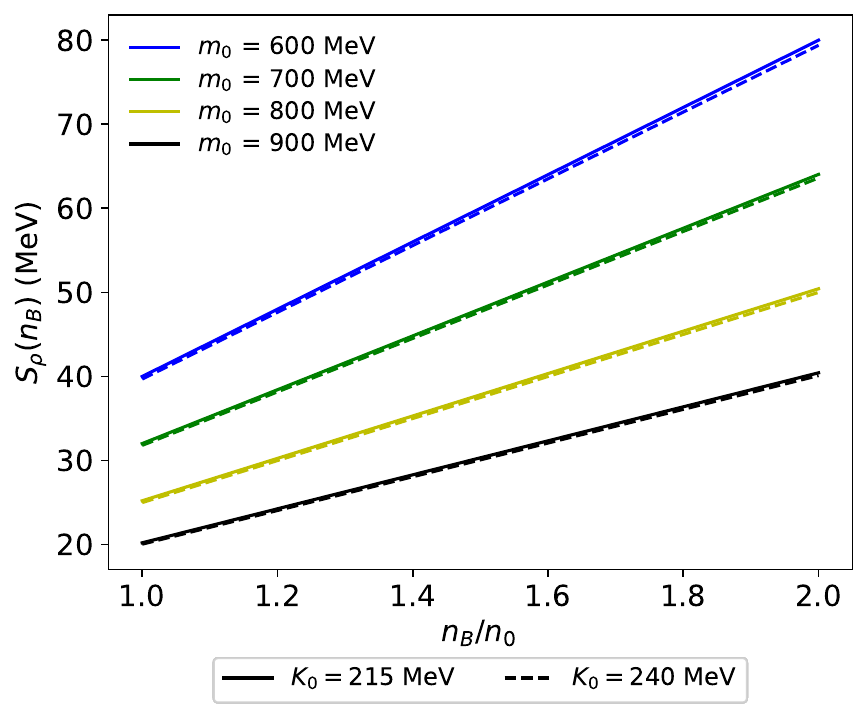}
\caption{$\rho$ meson contribution $S_\rho(n_B)$ in $a_0$ model for $m_0 = 600$-$900$ MeV and $\lambda_6'=0$. Solid curves represent $S_\rho(n_B)$ with $K_0 = 215$ MeV, while dashed curves represent $S_\rho(n_B)$ with $K_0 = 240$ MeV.\label{Srho without VM}}
\end{figure}

The $\rho$ meson contribution is given by 
\begin{equation}
\begin{aligned}
    S_\rho(n_B)\equiv  \frac{n_B}{2}\frac{(g_{\rho NN}/2)^2}{m_{\rho}^2 } \ ,
    \label{SrhoBa}
\end{aligned}
\end{equation} 
which is in the same form as Eq.~(\ref{Srho}). However, the $\rho$ meson coupling is much larger in the present model due to the attractive interaction of $a_0$(980). Figure~\ref{Srho without VM} shows the $S_\rho$ as a function of density. We observe that the value of $S_\rho$ is very large compared with the value in Figure~\ref{Sbrho without VM}.

\begin{figure}[H]
\centering
\includegraphics[width=10.5 cm]{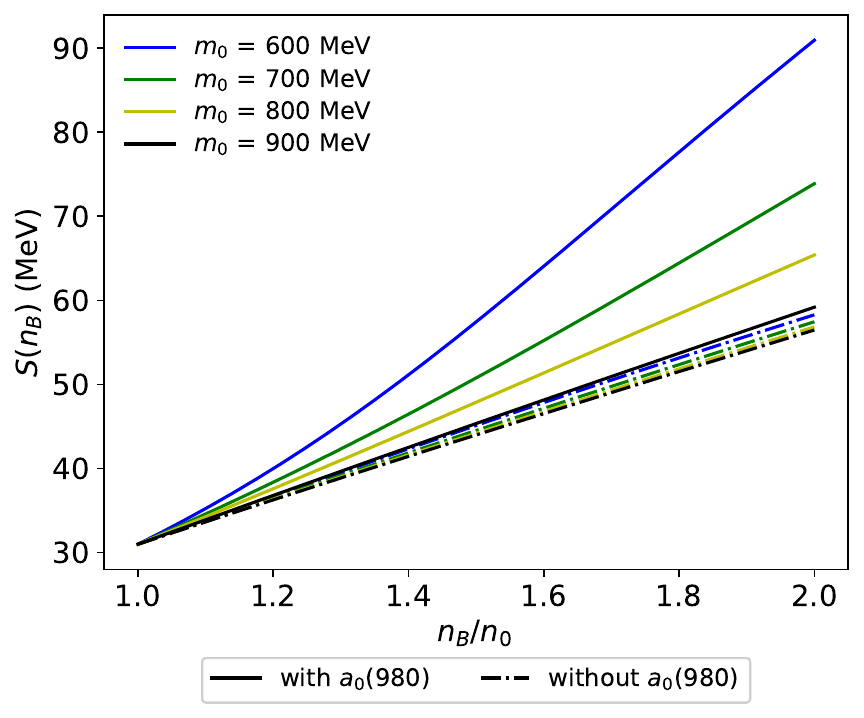}
\caption{Symmetry energy $S(n_B)$ for $m_0=600$-$900$\,MeV, $K_0 =215$ MeV, and $\lambda_6'=0$. Solid curves represent the $S(n_B)$ of the model including $a_0(980)$ with $\lambda_{6}'$=0, while the dash-dot curves show the results of the model without $a_0(980)$.\label{Scompare without VM}}
\end{figure}

Based on the above properties of three contributions,
the symmetry energy can be understood as a result of the competition between the repulsive $\rho$ meson interaction and the attractive $a_0$(980) interaction, in addition to the kinetic contribution from the nucleons. On the other hand, in the model without $a_0$ meson, only repulsive contributions exist. Since the symmetry energy at the saturation density is fixed as $S_0=31$\,MeV in the both  models with and without $a_0(980)$ meson, the $\rho$ meson coupling $g_{\rho NN}$ is strengthened by the existence of the attractive $a_0(980)$ contribution in the model with $a_0$ comparing to the model without $a_0$. Actually, it is clear from Tables~\ref{PD1215} and \ref{tab:gr 215} that $g_{\rho NN}$ is larger in the $a_0$ model than in the no-$a_0$ model for a fixed $m_0$. Figure~\ref{Scompare without VM} shows the symmetry energy for $m_0=600$-$900$ MeV and $K_0 =215$ MeV.

The result of the model without $a_0$ meson is shown in dashed line. We observe that the symmetry energy is indeed stiffened by the existence of $a_0$(980) and the difference of the symmetry energy between the models is larger for smaller $m_0$. At $n_B = 2n_0$, the symmetry energy $S(2n_0)$ is enlarged by as large as $\sim 60 \%$ or more in the present model depending on the choice of input parameters.

\begin{figure}[htbp]
\centering
\includegraphics[width=10.5 cm]{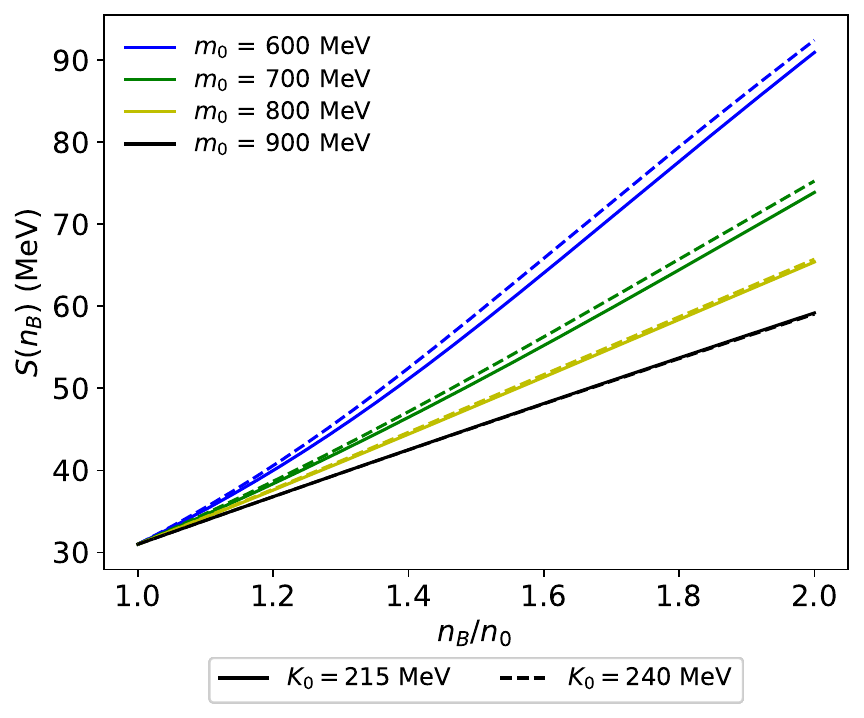}
\caption{Symmetry energy $S(n_B)$ for $m_0=600$-$900$\,MeV, $\lambda_6'=0$, with different choices of $K_0$ compared. Solid curves represent $S_(n_B)$ with $K_0 = 215$ MeV, while dashed curves represent $S(n_B)$ with $K_0 = 240$ MeV.\label{S without VM}}
\end{figure}

In Fig.~\ref{S without VM}, we compare the symmetry energy with different choices of $K_0$. We note that, similar to the previous model in Section~\ref{sec:PDM MF}, the symmetry energy is not sensitive to the value of $K_0$.

\begin{figure}[H]
\centering
\includegraphics[width=10.5 cm]{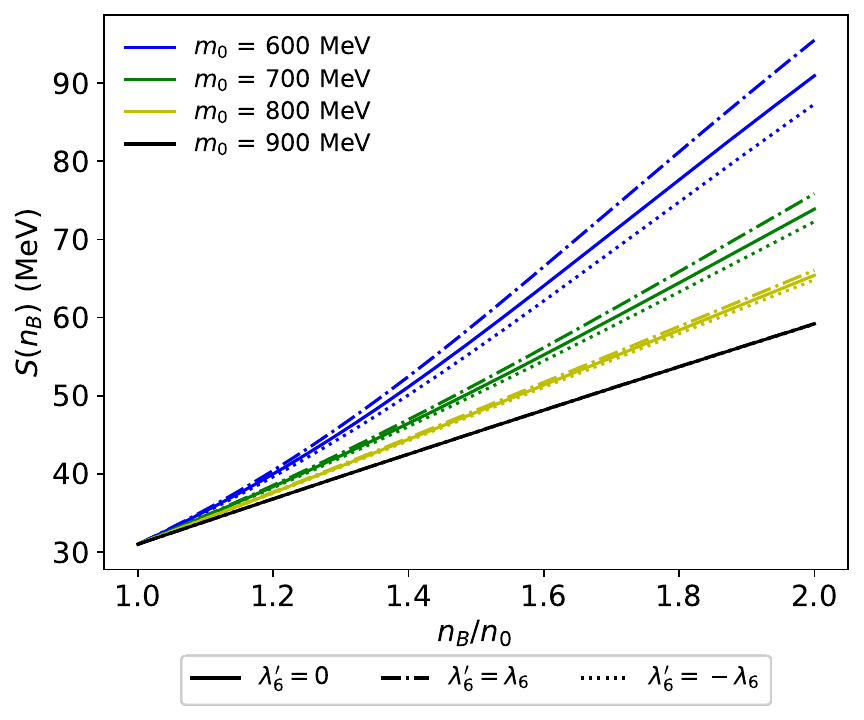}
\caption{Symmetry energy $S(n_B)$ for $m_0=600$-$900$\,MeV and $K_0 = 215$ MeV, with the effect of $\lambda_{6}' $ compared. Solid, dash-dot, and dotted curves show the $S(n_B)$ with $\lambda_{6}' = 0, \lambda_{6} $, and $-\lambda_{6}$, respectively. \label{Sl6p without VM}}
\end{figure}

In addition, we investigate the effect of higher-order terms in the large $N_c$ expansion for the six-point interaction on the symmetry energy
by taking $\lambda_6' = \pm \lambda_6$. The results of the symmetry energies with different values of $\lambda_{6}'$ are shown in Figure~\ref{Sl6p without VM}. We can see that the difference between the symmetry energies for models with the same $m_0$ is small, which indicates that the effect of $\lambda_6' $ on the symmetry energy is small. Notice also that the difference becomes smaller for larger $m_0$, due to a smaller $a_0$(980) effect.


\subsection{Symmetry energy of $a_0$ model with vector meson mixing}
\label{Sa0wVM}

As we see from the previous sections, PDM predicts a rather large slope parameter $L_0$ which seems not to be compatible to the recently accepted value of $L_0 = 57.7\pm 19$ MeV. In particular, the $a_0$ model in Section~\ref{sec:a0} predicts very large $L_0$ as well as the symmetry energy at density $n_B > n_0$. In order to soften the matter to reproduce the accepted value of $L_0$, we include the $\omega$-$\rho$ vector mixing, the $\lambda_{\omega \rho}$ term to effectively reduce the stiffness of the matter in our model. In this section, we study the symmetry energy with vector meson mixing interaction.

In the current model, the symmetry energy $S(n_B)$ for a given density $n_B$ is expressed as follows: 
\begin{equation}
\begin{aligned}
    S(n_B) & {} = \frac{n_B}{8} 
    \frac{\partial \mu_I}{\partial n_I} \Bigr|_{\substack{n_I=0}}\\
    & =  \frac{(k^*_+)^2}{6 \mu^*_+}   + \frac{n_B}{2}\frac{(g_{\rho NN}/2)^2}{m_{\rho}^2 + (2 \lambda_{\omega \rho}g_{\omega NN}^4 g_{\rho NN}^2 {{n_B^2}/{m_{\omega}^4}})} -  \frac{n_B}{4} \frac{m^*_+}{\mu^*_+}\frac{\partial m^*_{+n}}{\partial  n_I}\Bigr|_{\substack{n_I=0}} \ .
    \label{SymmetryE}
\end{aligned}
\end{equation} 
Similar to Eq.~(\ref{SrhoBawithoutVM}), the symmetry energy is divided into sum of three contributions: the nucleon contribution $S_N(n_B)$, $\rho$ meson contribution $S_\rho(n_B)$, and $a_0$ meson contribution $S_{a_0}(n_B)$. Notably,  the nucleon contribution and $a_0$ meson contribution are unaffected by the vector meson mixing since their related parameters are determined from symmetric matter properties. Therefore, the results of $S_N(n_B)$ and $S_{a_0}(n_B)$ are the same as given in Figs.~\ref{SbN without VM} and \ref{Sa without VM}.

\begin{figure}[htbp]
\centering
\includegraphics[width=10.5 cm]{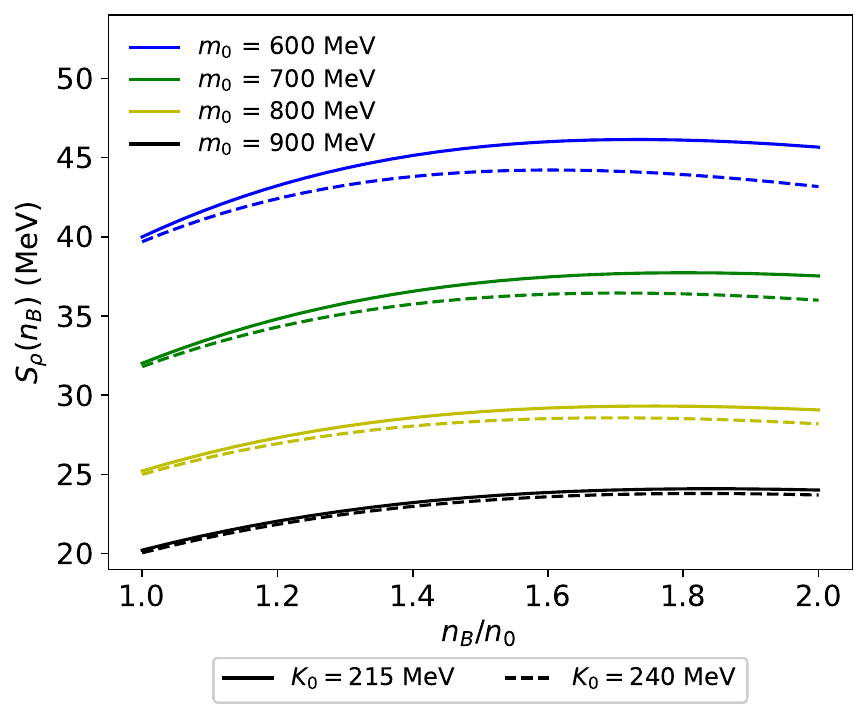}
\caption{$\rho$ meson contribution $S_\rho(n_B)$ in $a_0$ model for $m_0 = 600$-$900$ MeV, $\lambda_6'=0$, and $L_0 = 60$\,MeV. Solid curves represent $S_\rho(n_B)$ with $K_0 = 215$\,MeV, while dashed curves represent $S_\rho(n_B)$ with $K_0 = 240$\,MeV. \label{Srho with VM}}
\end{figure}

On the other hand, the $\rho$ meson contribution receives a correction from the vector meson mixing interaction as
\begin{equation}
\begin{aligned}
    S_{\rho}(n_B) \equiv \frac{n_B}{2}\Big[  \frac{(g_{\rho NN}/2)^2}{m_{\rho}^2 + (2 \lambda_{\omega \rho}g_{\omega NN}^4 g_{\rho NN}^2 n_B^2/m_{\omega}^4)} \Big],
    \label{Srho}
\end{aligned}
\end{equation}
where the $\rho$ meson appears to have an effective mass $({m^*_{\rho}})^2 = m_{\rho}^2 + (2 \lambda_{\omega \rho}g_{\omega NN}^4 g_{\rho NN}^2 n_B^2/m_{\omega}^4)$ exhibiting density-dependence. 
We note that the $\omega$ meson influences the symmetry energy through $2 \lambda_{\omega \rho}g_{\omega NN}^4 g_{\rho NN}^2 {{n_B^2}/{m_{\omega}^4}}$ in the denominator. Given that $\lambda_{\omega \rho}>0$ as shown in the previous section, the $\omega$-$\rho$ mixing term always reduces the symmetry energy. The density dependence of $S_{\rho}$ is illustrated in Figure~\ref{Srho with VM}. Furthermore, it is observed that $S_{\rho}$ increases with rising $n_B$ in the low-density region but decreases in the high-density region. This is understood as follows: in the low-density region where $m_\rho^2 \gg 2 \lambda_{\omega \rho}g_{\omega NN}^4 g_{\rho NN}^2 {{n_B^2}/{m_{\omega}^4}}$, the density dependence of $S(n_B)$ is primarily determined by the pre-factor $n_B$. In the high density region, on the other hand, the denominator is dominated by $2 \lambda_{\omega \rho}g_{\omega NN}^4 g_{\rho NN}^2 {{n_B^2}/{m_{\omega}^4}}$, which leads to $S_\rho(n_B) \propto 1/ n_B$. As a result, the behaviour of $S_\rho$ smoothly transforms from $\sim n_B \rightarrow \sim 1/ n_B$.

\begin{figure}[htbp]
\centering
\includegraphics[width=10.5 cm]{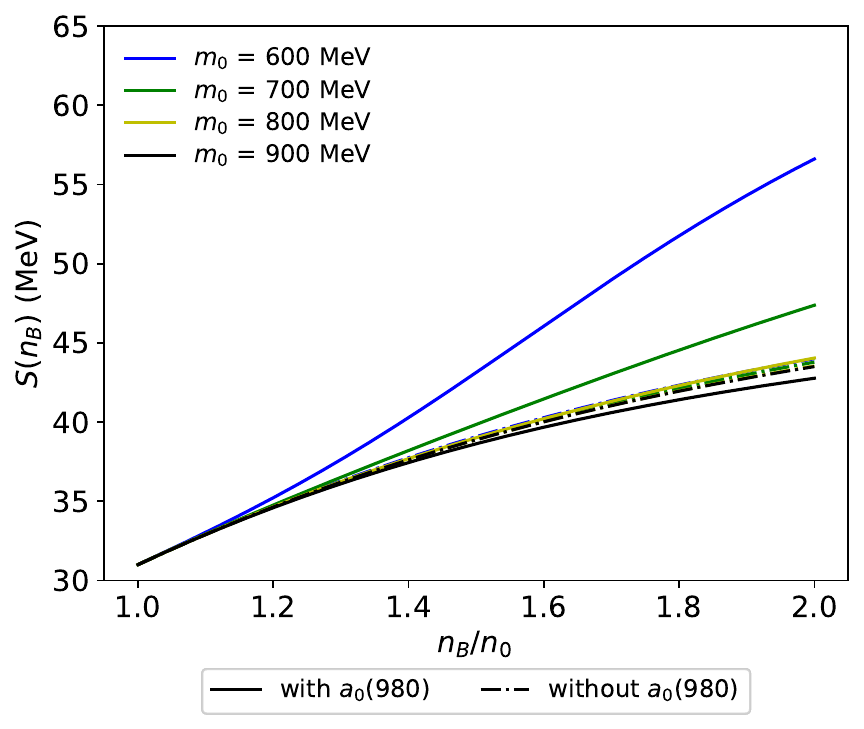}
\caption{Symmetry energy $S(n_B)$ in $a_0$ model with vector meson mixing for $m_0 = 600$-$900$\,MeV, $\lambda_6'=0$, and $L_0 = 60$\,MeV. Solid curves represent the $S(n_B)$ of the model including $a_0(980)$ with $\lambda_{6}'$=0, while the dash-dot curves show the results of the model without $a_0(980)$. \label{Scompare with VM L0060}}
\end{figure}
Figure~\ref{Scompare with VM L0060} shows the symmetry energy for $m_0=600$-$900$\,MeV and $L_0=60$\,MeV. For comparison, we also show the results of the no-$a_0$ model with vector meson mixing by dashed curves, as retrieved from Ref.~\cite{PhysRevC.108.055206}. 
We note that the density dependence is modified by the vector meson mixing interaction, where the slope of the symmetry energy is reduced in the high density region. We also observe that, in most cases, the symmetry energy is stiffened by the existence of $a_0$(980) and the difference of the symmetry energy between the models is larger for smaller $m_0$. In the case of large $m_0$ such as $m_0=900$ MeV where the $a_0$(980) meson effect is small, the softening effect of $\lambda_{\omega \rho}$ term overrides the stiffening effect from the $a_0$(980) meson. As a result, the symmetry energy $S(n_B)$ is reduced even after the inclusion of $a_0$ meson. A similar reduction of the symmetry energy in the intermediate density region was also reported in  Ref.~\cite{Li_2022} which includes both the scalar meson mixing and the vector meson mixing interactions in an RMF model with the presence of isovector-scalar meson.

% \begin{figure}[H]
% \begin{adjustwidth}{-\extralength}{0cm}
% %\centering
% \raggedleft
% \includegraphics[width=15.5cm]{Symmetry energy results/Scomparem0v with VM.pdf}
% \end{adjustwidth}
% \caption{$m_0$ dependence of thesymmetry energy $S(n_B)$ in $a_0$ model with vector meson mixing for $\lambda_6'=0$, $K_0 = 215$ MeV, and $L_0=40-80$ MeV.\label{ScompareL0v with VM}}
% \end{figure}  

% \kong{
% In addition, we investigate the $m_0$ dependence of the symmetry energy $S(n_B)$. Figure~\ref{ScompareL0v with VM} shows the comparison of symmetry energy with different $m_0$ for $L_0= 40-80$ MeV. We observe that smaller $m_0$ corresponding to a larger symmetry energy at density $n_B > n_0$. Moreover, the symmetry energy is increased more rapidly as $m_0$ becomes smaller at a fixed density $n_B > n_0$. 
% \hara{
% Also, we can see that larger $L_0$ leads to larger symmetry energy for fixed $m_0$, as we expected.
% [!We should put this kind of argument here to provide reasoning to show 5 plots. !]
% }

% [The last statement only true when density $n_B \lessapprox  2-4 n_0$ , in the case of model WITH vector mixing]

% [While for the a0 model WITHOUT vector mixing, the statement is always true (From my calculation, at least true for $n_B < 8 n_0$).]}

\begin{figure}[H]
\centering
\includegraphics[width=10.5 cm]{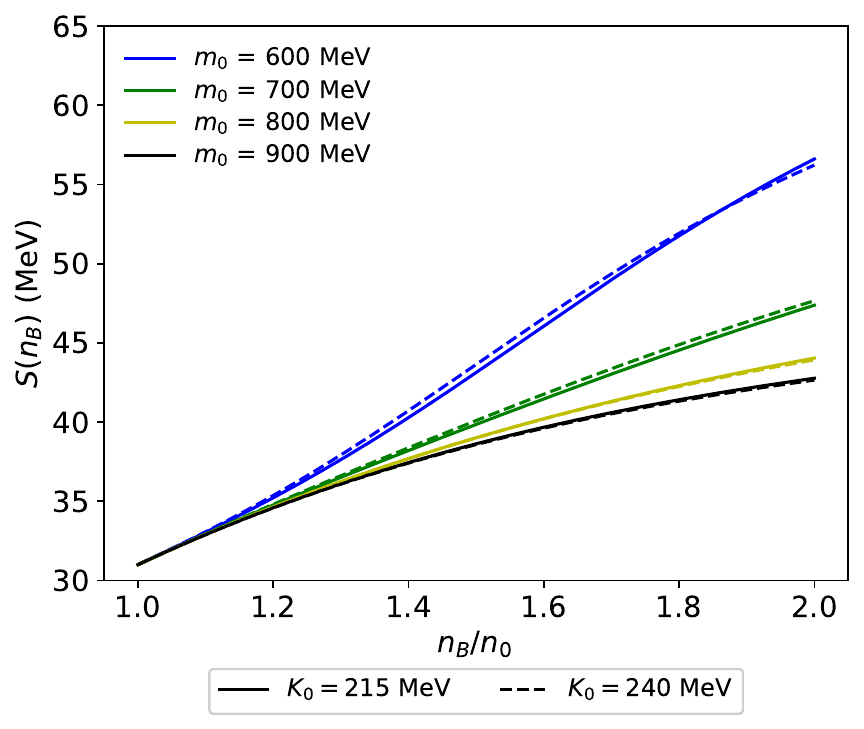}
\caption{$K_0$ dependence of the symmetry energy $S(n_B)$ in the $a_0$ model with vector meson mixing for $\lambda_6'=0$, and $L_0 = 60$\,MeV. Solid curves represent $S_\rho(n_B)$ with $K_0 = 215$\,MeV, while dashed curves represent $S_\rho(n_B)$ with $K_0 = 240$\,MeV. \label{ScompareK0 with VM}}
\end{figure}

In Fig.~\ref{ScompareK0 with VM}, we study the $K_0$ dependence of the symmetry energy. Similar to the models introduced in the previous sections, $K_0$ has very little effect to the symmetry energy.

\begin{figure}[H]
\centering
\includegraphics[width=10.5 cm]{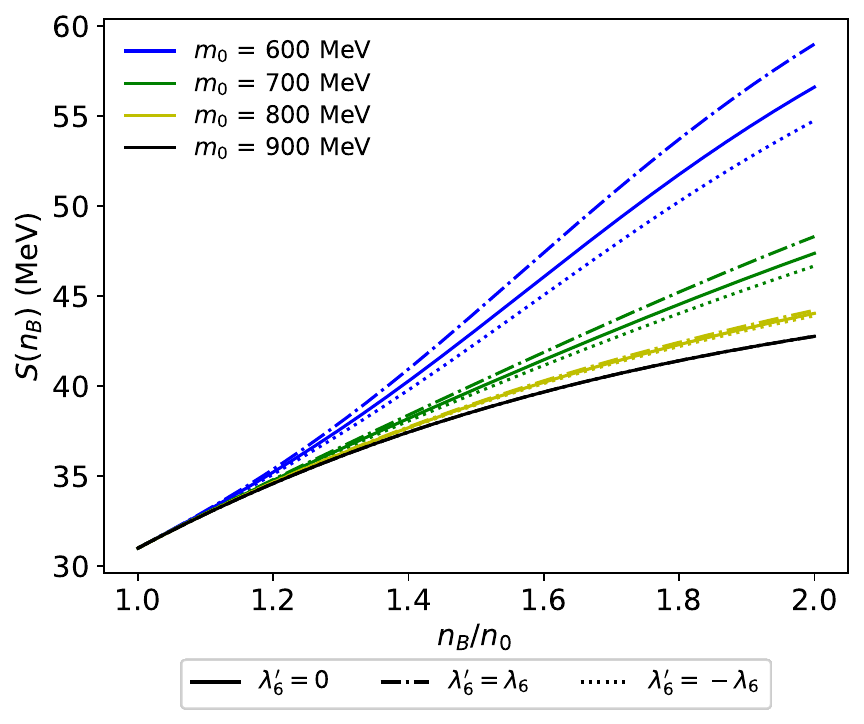}
\caption{Symmetry energy $S(n_B)$ for $m_0=600$-$900$\,MeV, $K_0 = 215$\,MeV, and $L_0 =60$\,MeV with the effect of $\lambda_{6}' $ compared. Solid, dash-dot, and dotted curves show the $S(n_B)$ with $\lambda_{6}' = 0, \pm \lambda_{6} $. \label{Sl6p with VM}}
\end{figure}

Finally, we compare the symmetry energy in the models with different $\lambda_6'$ in Fig.~\ref{Sl6p with VM}. As expected, the effect to symmetry energy is smaller than the effect of $m_0$, because $\lambda_6'$ interactions are of sub-leading order in large $N_c$ expansion.

%%%%%%%%%%%%%%%%%%%%%%%%%%%%%%%%%%%%%%%%%%
\section{Summary}\label{sec:summary}

In this review, we summarized the recent studies on infinite nuclear matter and finite nuclei based on 
parity doublet models (PDMs).
We first introduced a 
PDM which is 
constructed in Ref.~\cite{Motohiro:2015taa}. Under the mean field approximation, the nuclear properties such as slope parameter and symmetry energy were 
computed. In particular, we observed that the slope parameter is relatively larger than the recently accepted value $L_0 = 57.7 \pm 19 $\,MeV. We also investigated the impact of the value of $K_0$ to the model. We found
that the value of $K_0$ has little impact to the matter properties such as the symmetry energy and other model parameters.

We then considered the properties of some stable nuclei in the mean
field approximation to pin down the value of the chiral invariant mass preferred by the nuclear binding energies and
charge radii. We found that our results are closest to the experiments when we take $m_0 = 700$ MeV.  We also calculated the neutron and proton masses in a nucleus and observed, as expected, that the neutron-proton mass difference becomes larger in a isospin asymmetric nucleus.

Then, we introduced an extended PDM which incorporates the isovector scalar meson $a_0(980)$ into the matter. The isovector scalar meson provides the attractive force in the isovector channel and is 
important in the
asymmetric matter. We found that the inclusion of $a_0(980)$ has a very strong influence on the symmetry energy and slope parameter. We observe that the symmetry energy at densities $n_B > n_0$ is largely 
enhanced by the existence of $a_0(980)$. By analyzing the different contributions to the symmetry energy, we 
concluded that this enhancement is originated from the strengthening of the $\rho$ meson coupling $g_{\rho NN}$. The $a_0(980)$ meson generates the attaractive force in the isovector channel, which requires  the replulsive force from $\rho$ meson to be larger for reproducing the saturation property. As a result, a larger repulsive $\rho$ interaction increases the symmetry energy at densities $n_B > n_0$. However, we also observed that this stiffening of nuclear matter produces a large slope parameter that is much larger than the recently accepted value suggested by other studies. Therefore, we introduced the $\omega$-$\rho$ mixing interaction to reduce the slope parameter in the model. It was found that the symmetry energy at density $n_B > n_0$ is reduced after the inclusion of $\omega$-$\rho$ mixing interaction. Furthermore, we observed that the $\omega$-$\rho$ mixing interaction 
modifies the density dependence of the symmetry energy at density $n_B > n_0$.

We expect that future experiments on the study of symmetry energy at higher densities will provide further constraints on the chiral invariant mass of the nucleon. We also expect that $a_0(980)$ will have a significant influence on asymmetric nuclei. It would be interesting to study finite nuclei using the extended PDM including the $a_0(980)$ meson, which may give new information to constraints on the chiral invariant mass of the nucleon and the behaviour of nucleon mass in dense matter. We leave this as future project.

\acknowledgments{
We thank Jie Meng for providing us the RCHB code. 
This work was supported in part by the Institute for Basic Science (IBS-R031-D1) (YK), 
JSPS KAKENHI Grant Nos.~20K03927, 23H05439
and 24K07045 (MH). 
}

\begin{adjustwidth}{-\extralength}{0cm}

\reftitle{References}


\bibliography{PDM-a0-refs}

\begin{thebibliography}{999}

\bibitem[Aarts et~al.(2015)Aarts, Allton, Hands, J\"ager, Praki, and
  Skullerud]{PhysRevD.92.014503}
Aarts, G.; Allton, C.; Hands, S.; J\"ager, B.; Praki, C.; Skullerud, J.I.
\newblock Nucleons and parity doubling across the deconfinement transition.
\newblock {\em Phys. Rev. D} {\bf 2015}, {\em 92},~014503.
\newblock {\url{https://doi.org/10.1103/PhysRevD.92.014503}}.

\bibitem[{Aarts} et~al.(2017){Aarts}, {Allton}, {De Boni}, {Hands},
  {J{\"a}ger}, {Praki}, and {Skullerud}]{2017JHEP...06..034A}
{Aarts}, G.; {Allton}, C.; {De Boni}, D.; {Hands}, S.; {J{\"a}ger}, B.;
  {Praki}, C.; {Skullerud}, J.I.
\newblock {Light baryons below and above the deconfinement transition: medium
  effects and parity doubling}.
\newblock {\em Journal of High Energy Physics} {\bf 2017}, {\em 2017},~34,
  \href{http://xxx.lanl.gov/abs/1703.09246}{{\normalfont
  [arXiv:hep-lat/1703.09246]}}.
\newblock {\url{https://doi.org/10.1007/JHEP06(2017)034}}.

\bibitem[DeTar and Kunihiro(1989)]{PhysRevD.39.2805}
DeTar, C.; Kunihiro, T.
\newblock Linear sigma model with parity doubling.
\newblock {\em Phys. Rev. D} {\bf 1989}, {\em 39},~2805--2808.
\newblock {\url{https://doi.org/10.1103/PhysRevD.39.2805}}.

\bibitem[Kim and Lee(2022)]{PhysRevD.105.014014}
Kim, J.; Lee, S.H.
\newblock Masses of hadrons in the chiral symmetry restored vacuum.
\newblock {\em Phys. Rev. D} {\bf 2022}, {\em 105},~014014.
\newblock {\url{https://doi.org/10.1103/PhysRevD.105.014014}}.

\bibitem[Jido et~al.(2001)Jido, Oka, and Hosaka]{10.1143/PTP.106.873}
Jido, D.; Oka, M.; Hosaka, A.
\newblock {Chiral Symmetry of Baryons}.
\newblock {\em Progress of Theoretical Physics} {\bf 2001}, {\em
  106},~873--908,
  \href{http://xxx.lanl.gov/abs/https://academic.oup.com/ptp/article-pdf/106/5/873/5373808/106-5-873.pdf}{{\normalfont
  [https://academic.oup.com/ptp/article-pdf/106/5/873/5373808/106-5-873.pdf]}}.
\newblock {\url{https://doi.org/10.1143/PTP.106.873}}.

\bibitem[Yamazaki and Harada(2019)]{Yamazaki:2018stk}
Yamazaki, T.; Harada, M.
\newblock {Chiral partner structure of light nucleons in an extended parity
  doublet model}.
\newblock {\em Phys. Rev. D} {\bf 2019}, {\em 99},~034012,
  \href{http://xxx.lanl.gov/abs/1809.02359}{{\normalfont
  [arXiv:hep-ph/1809.02359]}}.
\newblock {\url{https://doi.org/10.1103/PhysRevD.99.034012}}.

\bibitem[Hatsuda and Prakash(1989)]{Hatsuda:1988mv}
Hatsuda, T.; Prakash, M.
\newblock {Parity Doubling of the Nucleon and First Order Chiral Transition in
  Dense Matter}.
\newblock {\em Phys. Lett. B} {\bf 1989}, {\em 224},~11--15.
\newblock {\url{https://doi.org/10.1016/0370-2693(89)91040-X}}.

\bibitem[Zschiesche et~al.(2007)Zschiesche, Tolos, Schaffner-Bielich, and
  Pisarski]{Zschiesche:2006zj}
Zschiesche, D.; Tolos, L.; Schaffner-Bielich, J.; Pisarski, R.D.
\newblock {Cold, dense nuclear matter in a SU(2) parity doublet model}.
\newblock {\em Phys. Rev. C} {\bf 2007}, {\em 75},~055202,
  \href{http://xxx.lanl.gov/abs/nucl-th/0608044}{{\normalfont
  [nucl-th/0608044]}}.
\newblock {\url{https://doi.org/10.1103/PhysRevC.75.055202}}.

\bibitem[Dexheimer et~al.(2008{\natexlab{a}})Dexheimer, Schramm, and
  Zschiesche]{Dexheimer:2007tn}
Dexheimer, V.; Schramm, S.; Zschiesche, D.
\newblock {Nuclear matter and neutron stars in a parity doublet model}.
\newblock {\em Phys. Rev. C} {\bf 2008}, {\em 77},~025803,
  \href{http://xxx.lanl.gov/abs/0710.4192}{{\normalfont
  [arXiv:nucl-th/0710.4192]}}.
\newblock {\url{https://doi.org/10.1103/PhysRevC.77.025803}}.

\bibitem[Dexheimer et~al.(2008{\natexlab{b}})Dexheimer, Pagliara, Tolos,
  Schaffner-Bielich, and Schramm]{Dexheimer:2008cv}
Dexheimer, V.; Pagliara, G.; Tolos, L.; Schaffner-Bielich, J.; Schramm, S.
\newblock {Neutron stars within the SU(2) parity doublet model}.
\newblock {\em Eur. Phys. J. A} {\bf 2008}, {\em 38},~105--113,
  \href{http://xxx.lanl.gov/abs/0805.3301}{{\normalfont
  [arXiv:nucl-th/0805.3301]}}.
\newblock {\url{https://doi.org/10.1140/epja/i2008-10652-0}}.

\bibitem[Sasaki and Mishustin(2010)]{Sasaki:2010bp}
Sasaki, C.; Mishustin, I.
\newblock {Thermodynamics of dense hadronic matter in a parity doublet model}.
\newblock {\em Phys. Rev. C} {\bf 2010}, {\em 82},~035204,
  \href{http://xxx.lanl.gov/abs/1005.4811}{{\normalfont
  [arXiv:hep-ph/1005.4811]}}.
\newblock {\url{https://doi.org/10.1103/PhysRevC.82.035204}}.

\bibitem[Sasaki et~al.(2011)Sasaki, Lee, Paeng, and Rho]{Sasaki:2011ff}
Sasaki, C.; Lee, H.K.; Paeng, W.G.; Rho, M.
\newblock {Conformal anomaly and the vector coupling in dense matter}.
\newblock {\em Phys. Rev. D} {\bf 2011}, {\em 84},~034011,
  \href{http://xxx.lanl.gov/abs/1103.0184}{{\normalfont
  [arXiv:hep-ph/1103.0184]}}.
\newblock {\url{https://doi.org/10.1103/PhysRevD.84.034011}}.

\bibitem[Gallas et~al.(2011)Gallas, Giacosa, and Pagliara]{Gallas:2011qp}
Gallas, S.; Giacosa, F.; Pagliara, G.
\newblock {Nuclear matter within a dilatation-invariant parity doublet model:
  The role of the tetraquark at nonzero density}.
\newblock {\em Nucl. Phys. A} {\bf 2011}, {\em 872},~13--24,
  \href{http://xxx.lanl.gov/abs/1105.5003}{{\normalfont
  [arXiv:hep-ph/1105.5003]}}.
\newblock {\url{https://doi.org/10.1016/j.nuclphysa.2011.09.008}}.

\bibitem[Paeng et~al.(2012)Paeng, Lee, Rho, and Sasaki]{Paeng:2011hy}
Paeng, W.G.; Lee, H.K.; Rho, M.; Sasaki, C.
\newblock {Dilaton-Limit Fixed Point in Hidden Local Symmetric Parity Doublet
  Model}.
\newblock {\em Phys. Rev. D} {\bf 2012}, {\em 85},~054022,
  \href{http://xxx.lanl.gov/abs/1109.5431}{{\normalfont
  [arXiv:hep-ph/1109.5431]}}.
\newblock {\url{https://doi.org/10.1103/PhysRevD.85.054022}}.

\bibitem[Steinheimer et~al.(2011)Steinheimer, Schramm, and
  Stocker]{Steinheimer:2011ea}
Steinheimer, J.; Schramm, S.; Stocker, H.
\newblock {The hadronic SU(3) Parity Doublet Model for Dense Matter, its
  extension to quarks and the strange equation of state}.
\newblock {\em Phys. Rev. C} {\bf 2011}, {\em 84},~045208,
  \href{http://xxx.lanl.gov/abs/1108.2596}{{\normalfont
  [arXiv:hep-ph/1108.2596]}}.
\newblock {\url{https://doi.org/10.1103/PhysRevC.84.045208}}.

\bibitem[Dexheimer et~al.(2013)Dexheimer, Steinheimer, Negreiros, and
  Schramm]{Dexheimer:2012eu}
Dexheimer, V.; Steinheimer, J.; Negreiros, R.; Schramm, S.
\newblock {Hybrid Stars in an SU(3) parity doublet model}.
\newblock {\em Phys. Rev. C} {\bf 2013}, {\em 87},~015804,
  \href{http://xxx.lanl.gov/abs/1206.3086}{{\normalfont
  [arXiv:astro-ph.HE/1206.3086]}}.
\newblock {\url{https://doi.org/10.1103/PhysRevC.87.015804}}.

\bibitem[Paeng et~al.(2013)Paeng, Lee, Rho, and Sasaki]{Paeng:2013xya}
Paeng, W.G.; Lee, H.K.; Rho, M.; Sasaki, C.
\newblock {Interplay between $\omega$-nucleon interaction and nucleon mass in
  dense baryonic matter}.
\newblock {\em Phys. Rev. D} {\bf 2013}, {\em 88},~105019,
  \href{http://xxx.lanl.gov/abs/1303.2898}{{\normalfont
  [arXiv:nucl-th/1303.2898]}}.
\newblock {\url{https://doi.org/10.1103/PhysRevD.88.105019}}.

\bibitem[Benic et~al.(2015)Benic, Mishustin, and Sasaki]{Benic:2015pia}
Benic, S.; Mishustin, I.; Sasaki, C.
\newblock {Effective model for the QCD phase transitions at finite baryon
  density}.
\newblock {\em Phys. Rev. D} {\bf 2015}, {\em 91},~125034,
  \href{http://xxx.lanl.gov/abs/1502.05969}{{\normalfont
  [arXiv:hep-ph/1502.05969]}}.
\newblock {\url{https://doi.org/10.1103/PhysRevD.91.125034}}.

\bibitem[Motohiro et~al.(2015)Motohiro, Kim, and Harada]{Motohiro:2015taa}
Motohiro, Y.; Kim, Y.; Harada, M.
\newblock {Asymmetric nuclear matter in a parity doublet model with hidden
  local symmetry}.
\newblock {\em Phys. Rev. C} {\bf 2015}, {\em 92},~025201,
  \href{http://xxx.lanl.gov/abs/1505.00988}{{\normalfont
  [arXiv:nucl-th/1505.00988]}}.
\newblock [Erratum: Phys.Rev.C 95, 059903 (2017)],
  {\url{https://doi.org/10.1103/PhysRevC.92.025201}}.

\bibitem[Mukherjee et~al.(2017)Mukherjee, Steinheimer, and
  Schramm]{Mukherjee:2016nhb}
Mukherjee, A.; Steinheimer, J.; Schramm, S.
\newblock {Higher-order baryon number susceptibilities: Interplay between the
  chiral and the nuclear liquid-gas transitions}.
\newblock {\em Phys. Rev. C} {\bf 2017}, {\em 96},~025205,
  \href{http://xxx.lanl.gov/abs/1611.10144}{{\normalfont
  [arXiv:nucl-th/1611.10144]}}.
\newblock {\url{https://doi.org/10.1103/PhysRevC.96.025205}}.

\bibitem[Suenaga(2018)]{Suenaga:2017wbb}
Suenaga, D.
\newblock {Examination of $N^*(1535)$ as a probe to observe the partial
  restoration of chiral symmetry in nuclear matter}.
\newblock {\em Phys. Rev. C} {\bf 2018}, {\em 97},~045203,
  \href{http://xxx.lanl.gov/abs/1704.03630}{{\normalfont
  [arXiv:nucl-th/1704.03630]}}.
\newblock {\url{https://doi.org/10.1103/PhysRevC.97.045203}}.

\bibitem[Takeda et~al.(2018)Takeda, Kim, and Harada]{Takeda:2017mrm}
Takeda, Y.; Kim, Y.; Harada, M.
\newblock {Catalysis of partial chiral symmetry restoration by $\Delta$
  matter}.
\newblock {\em Phys. Rev. C} {\bf 2018}, {\em 97},~065202,
  \href{http://xxx.lanl.gov/abs/1704.04357}{{\normalfont
  [arXiv:nucl-th/1704.04357]}}.
\newblock {\url{https://doi.org/10.1103/PhysRevC.97.065202}}.

\bibitem[Mukherjee et~al.(2017)Mukherjee, Schramm, Steinheimer, and
  Dexheimer]{Mukherjee:2017jzi}
Mukherjee, A.; Schramm, S.; Steinheimer, J.; Dexheimer, V.
\newblock {The application of the Quark-Hadron Chiral Parity-Doublet Model to
  neutron star matter}.
\newblock {\em Astron. Astrophys.} {\bf 2017}, {\em 608},~A110,
  \href{http://xxx.lanl.gov/abs/1706.09191}{{\normalfont
  [arXiv:nucl-th/1706.09191]}}.
\newblock {\url{https://doi.org/10.1051/0004-6361/201731505}}.

\bibitem[Paeng et~al.(2017)Paeng, Kuo, Lee, Ma, and Rho]{Paeng:2017qvp}
Paeng, W.G.; Kuo, T.T.S.; Lee, H.K.; Ma, Y.L.; Rho, M.
\newblock {Scale-invariant hidden local symmetry, topology change, and dense
  baryonic matter. II.}
\newblock {\em Phys. Rev. D} {\bf 2017}, {\em 96},~014031,
  \href{http://xxx.lanl.gov/abs/1704.02775}{{\normalfont
  [arXiv:nucl-th/1704.02775]}}.
\newblock {\url{https://doi.org/10.1103/PhysRevD.96.014031}}.

\bibitem[Marczenko and Sasaki(2018)]{Marczenko:2017huu}
Marczenko, M.; Sasaki, C.
\newblock {Net-baryon number fluctuations in the Hybrid Quark-Meson-Nucleon
  model at finite density}.
\newblock {\em Phys. Rev. D} {\bf 2018}, {\em 97},~036011,
  \href{http://xxx.lanl.gov/abs/1711.05521}{{\normalfont
  [arXiv:hep-ph/1711.05521]}}.
\newblock {\url{https://doi.org/10.1103/PhysRevD.97.036011}}.

\bibitem[Abuki et~al.(2018)Abuki, Takeda, and Harada]{Abuki:2018ijb}
Abuki, H.; Takeda, Y.; Harada, M.
\newblock {Dual chiral density waves in nuclear matter}.
\newblock {\em Epj Web Conf.} {\bf 2018}, {\em 192},~00020,
  \href{http://xxx.lanl.gov/abs/1809.06485}{{\normalfont
  [arXiv:hep-ph/1809.06485]}}.
\newblock {\url{https://doi.org/10.1051/epjconf/201819200020}}.

\bibitem[Marczenko et~al.(2018)Marczenko, Blaschke, Redlich, and
  Sasaki]{Marczenko:2018jui}
Marczenko, M.; Blaschke, D.; Redlich, K.; Sasaki, C.
\newblock {Chiral symmetry restoration by parity doubling and the structure of
  neutron stars}.
\newblock {\em Phys. Rev. D} {\bf 2018}, {\em 98},~103021,
  \href{http://xxx.lanl.gov/abs/1805.06886}{{\normalfont
  [arXiv:nucl-th/1805.06886]}}.
\newblock {\url{https://doi.org/10.1103/PhysRevD.98.103021}}.

\bibitem[Marczenko et~al.(2019)Marczenko, Blaschke, Redlich, and
  Sasaki]{Marczenko:2019trv}
Marczenko, M.; Blaschke, D.; Redlich, K.; Sasaki, C.
\newblock {Parity Doubling and the Dense Matter Phase Diagram under Constraints
  from Multi-Messenger Astronomy}.
\newblock {\em Universe} {\bf 2019}, {\em 5},~180,
  \href{http://xxx.lanl.gov/abs/1905.04974}{{\normalfont
  [arXiv:nucl-th/1905.04974]}}.
\newblock {\url{https://doi.org/10.3390/universe5080180}}.

\bibitem[Yamazaki and Harada(2019)]{Yamazaki:2019tuo}
Yamazaki, T.; Harada, M.
\newblock {Constraint to chiral invariant masses of nucleons from GW170817 in
  an extended parity doublet model}.
\newblock {\em Phys. Rev. C} {\bf 2019}, {\em 100},~025205,
  \href{http://xxx.lanl.gov/abs/1901.02167}{{\normalfont
  [arXiv:nucl-th/1901.02167]}}.
\newblock {\url{https://doi.org/10.1103/PhysRevC.100.025205}}.

\bibitem[Harada and Yamazaki(2019)]{Harada:2019oaq}
Harada, M.; Yamazaki, T.
\newblock {Charmed Mesons in Nuclear Matter Based on Chiral Effective Models}.
\newblock {\em Jps Conf. Proc.} {\bf 2019}, {\em 26},~024001.
\newblock {\url{https://doi.org/10.7566/JPSCP.26.024001}}.

\bibitem[Marczenko et~al.(2020)Marczenko, Blaschke, Redlich, and
  Sasaki]{Marczenko:2020jma}
Marczenko, M.; Blaschke, D.; Redlich, K.; Sasaki, C.
\newblock {Toward a unified equation of state for multi-messenger astronomy}.
\newblock {\em Astron. Astrophys.} {\bf 2020}, {\em 643},~A82,
  \href{http://xxx.lanl.gov/abs/2004.09566}{{\normalfont
  [arXiv:astro-ph.HE/2004.09566]}}.
\newblock {\url{https://doi.org/10.1051/0004-6361/202038211}}.

\bibitem[Harada(2020)]{Harada:2020etl}
Harada, M.
\newblock {Dense nuclear matter based on a chiral model with parity doublet
  structure}.
\newblock In Proceedings of the {18th International Conference on Hadron
  Spectroscopy and Structure},  2020.
\newblock {\url{https://doi.org/10.1142/9789811219313_0113}}.

\bibitem[Minamikawa et~al.(2021)Minamikawa, Kojo, and
  Harada]{PhysRevC.103.045205}
Minamikawa, T.; Kojo, T.; Harada, M.
\newblock Quark-hadron crossover equations of state for neutron stars:
  Constraining the chiral invariant mass in a parity doublet model.
\newblock {\em Phys. Rev. C} {\bf 2021}, {\em 103},~045205.
\newblock {\url{https://doi.org/10.1103/PhysRevC.103.045205}}.

\bibitem[Marczenko et~al.(2022)Marczenko, Redlich, and
  Sasaki]{Marczenko:2021uaj}
Marczenko, M.; Redlich, K.; Sasaki, C.
\newblock {Reconciling Multi-messenger Constraints with Chiral Symmetry
  Restoration}.
\newblock {\em Astrophys. J. Lett.} {\bf 2022}, {\em 925},~L23,
  \href{http://xxx.lanl.gov/abs/2110.11056}{{\normalfont
  [arXiv:nucl-th/2110.11056]}}.
\newblock {\url{https://doi.org/10.3847/2041-8213/ac4b61}}.

\bibitem[Minamikawa et~al.(2021)Minamikawa, Kojo, and
  Harada]{PhysRevC.104.065201}
Minamikawa, T.; Kojo, T.; Harada, M.
\newblock Chiral condensates for neutron stars in hadron-quark crossover: From
  a parity doublet nucleon model to a Nambu--Jona-Lasinio quark model.
\newblock {\em Phys. Rev. C} {\bf 2021}, {\em 104},~065201.
\newblock {\url{https://doi.org/10.1103/PhysRevC.104.065201}}.

\bibitem[Marczenko et~al.(2022)Marczenko, Redlich, and
  Sasaki]{Marczenko:2022hyt}
Marczenko, M.; Redlich, K.; Sasaki, C.
\newblock {Chiral symmetry restoration and \ensuremath{\Delta} matter formation
  in neutron stars}.
\newblock {\em Phys. Rev. D} {\bf 2022}, {\em 105},~103009,
  \href{http://xxx.lanl.gov/abs/2203.00269}{{\normalfont
  [arXiv:nucl-th/2203.00269]}}.
\newblock {\url{https://doi.org/10.1103/PhysRevD.105.103009}}.

\bibitem[Gao et~al.(2022)Gao, Minamikawa, Kojo, and
  Harada]{PhysRevC.106.065205}
Gao, B.; Minamikawa, T.; Kojo, T.; Harada, M.
\newblock Impacts of the $\mathrm{U}{(1)}_{A}$ anomaly on nuclear and neutron
  star equation of state based on a parity doublet model.
\newblock {\em Phys. Rev. C} {\bf 2022}, {\em 106},~065205.
\newblock {\url{https://doi.org/10.1103/PhysRevC.106.065205}}.

\bibitem[Minamikawa et~al.(2023)Minamikawa, Gao, Kojo, and
  Harada]{Minamikawa:2023eky}
Minamikawa, T.; Gao, B.; Kojo, T.; Harada, M.
\newblock {Chiral restoration of nucleons in neutron star matter: studies based
  on a parity doublet model} {\bf 2023}.
\newblock  \href{http://xxx.lanl.gov/abs/2302.00825}{{\normalfont
  [arXiv:nucl-th/2302.00825]}}.

\bibitem[Marczenko et~al.(2023)Marczenko, Redlich, and
  Sasaki]{Marczenko:2023ohi}
Marczenko, M.; Redlich, K.; Sasaki, C.
\newblock {Fluctuations near the liquid-gas and chiral phase transitions in
  hadronic matter} {\bf 2023}.
\newblock  \href{http://xxx.lanl.gov/abs/2301.09866}{{\normalfont
  [arXiv:nucl-th/2301.09866]}}.

\bibitem[Baym et~al.(2018)Baym, Hatsuda, Kojo, Powell, Song, and
  Takatsuka]{Baym_2018}
Baym, G.; Hatsuda, T.; Kojo, T.; Powell, P.D.; Song, Y.; Takatsuka, T.
\newblock From hadrons to quarks in neutron stars: a review.
\newblock {\em Reports on Progress in Physics} {\bf 2018}, {\em 81},~056902.
\newblock {\url{https://doi.org/10.1088/1361-6633/aaae14}}.

\bibitem[Baym et~al.(2019)Baym, Furusawa, Hatsuda, Kojo, and
  Togashi]{Baym_2019}
Baym, G.; Furusawa, S.; Hatsuda, T.; Kojo, T.; Togashi, H.
\newblock New Neutron Star Equation of State with Quark--Hadron Crossover.
\newblock {\em The Astrophysical Journal} {\bf 2019}, {\em 885},~42.
\newblock {\url{https://doi.org/10.3847/1538-4357/ab441e}}.

\bibitem[Cromartie et~al.(2019)]{NANOGrav:2019jur}
Cromartie, H.T.;  et~al.
\newblock {Relativistic Shapiro delay measurements of an extremely massive
  millisecond pulsar}.
\newblock {\em Nature Astron.} {\bf 2019}, {\em 4},~72--76,
  \href{http://xxx.lanl.gov/abs/1904.06759}{{\normalfont
  [arXiv:astro-ph.HE/1904.06759]}}.
\newblock {\url{https://doi.org/10.1038/s41550-019-0880-2}}.

\bibitem[Abbott et~al.(2017{\natexlab{a}})]{LIGOScientific:2017vwq}
Abbott, B.P.;  et~al.
\newblock {GW170817: Observation of Gravitational Waves from a Binary Neutron
  Star Inspiral}.
\newblock {\em Phys. Rev. Lett.} {\bf 2017}, {\em 119},~161101,
  \href{http://xxx.lanl.gov/abs/1710.05832}{{\normalfont
  [arXiv:gr-qc/1710.05832]}}.
\newblock {\url{https://doi.org/10.1103/PhysRevLett.119.161101}}.

\bibitem[Abbott et~al.(2017{\natexlab{b}})]{LIGOScientific:2017ync}
Abbott, B.P.;  et~al.
\newblock {Multi-messenger Observations of a Binary Neutron Star Merger}.
\newblock {\em Astrophys. J. Lett.} {\bf 2017}, {\em 848},~L12,
  \href{http://xxx.lanl.gov/abs/1710.05833}{{\normalfont
  [arXiv:astro-ph.HE/1710.05833]}}.
\newblock {\url{https://doi.org/10.3847/2041-8213/aa91c9}}.

\bibitem[Abbott et~al.(2017{\natexlab{c}})]{TheLIGOScientific:2017qsa}
Abbott, B.;  et~al.
\newblock {GW170817: Observation of Gravitational Waves from a Binary Neutron
  Star Inspiral}.
\newblock {\em Phys. Rev. Lett.} {\bf 2017}, {\em 119},~161101,
  \href{http://xxx.lanl.gov/abs/1710.05832}{{\normalfont
  [arXiv:gr-qc/1710.05832]}}.
\newblock {\url{https://doi.org/10.1103/PhysRevLett.119.161101}}.

\bibitem[Abbott et~al.(2018)]{LIGOScientific:2018cki}
Abbott, B.P.;  et~al.
\newblock {GW170817: Measurements of neutron star radii and equation of state}.
\newblock {\em Phys. Rev. Lett.} {\bf 2018}, {\em 121},~161101,
  \href{http://xxx.lanl.gov/abs/1805.11581}{{\normalfont
  [arXiv:gr-qc/1805.11581]}}.
\newblock {\url{https://doi.org/10.1103/PhysRevLett.121.161101}}.

\bibitem[Miller et~al.(2019)]{Miller:2019cac}
Miller, M.;  et~al.
\newblock {PSR J0030+0451 Mass and Radius from $NICER$ Data and Implications
  for the Properties of Neutron Star Matter}.
\newblock {\em Astrophys. J. Lett.} {\bf 2019}, {\em 887},~L24,
  \href{http://xxx.lanl.gov/abs/1912.05705}{{\normalfont
  [arXiv:astro-ph.HE/1912.05705]}}.
\newblock {\url{https://doi.org/10.3847/2041-8213/ab50c5}}.

\bibitem[Riley et~al.(2019)]{Riley:2019yda}
Riley, T.E.;  et~al.
\newblock {A $NICER$ View of PSR J0030+0451: Millisecond Pulsar Parameter
  Estimation}.
\newblock {\em Astrophys. J. Lett.} {\bf 2019}, {\em 887},~L21,
  \href{http://xxx.lanl.gov/abs/1912.05702}{{\normalfont
  [arXiv:astro-ph.HE/1912.05702]}}.
\newblock {\url{https://doi.org/10.3847/2041-8213/ab481c}}.

\bibitem[Fonseca et~al.(2021)]{Fonseca:2021wxt}
Fonseca, E.;  et~al.
\newblock {Refined Mass and Geometric Measurements of the High-mass PSR
  J0740+6620}.
\newblock {\em Astrophys. J. Lett.} {\bf 2021}, {\em 915},~L12,
  \href{http://xxx.lanl.gov/abs/2104.00880}{{\normalfont
  [arXiv:astro-ph.HE/2104.00880]}}.
\newblock {\url{https://doi.org/10.3847/2041-8213/ac03b8}}.

\bibitem[De et~al.(2018)De, Finstad, Lattimer, Brown, Berger, and
  Biwer]{De:2018uhw}
De, S.; Finstad, D.; Lattimer, J.M.; Brown, D.A.; Berger, E.; Biwer, C.M.
\newblock {Tidal Deformabilities and Radii of Neutron Stars from the
  Observation of GW170817}.
\newblock {\em Phys. Rev. Lett.} {\bf 2018}, {\em 121},~091102,
  \href{http://xxx.lanl.gov/abs/1804.08583}{{\normalfont
  [arXiv:astro-ph.HE/1804.08583]}}.
\newblock [Erratum: Phys.Rev.Lett. 121, 259902 (2018)],
  {\url{https://doi.org/10.1103/PhysRevLett.121.091102}}.

\bibitem[Radice et~al.(2018)Radice, Perego, Zappa, and
  Bernuzzi]{Radice:2017lry}
Radice, D.; Perego, A.; Zappa, F.; Bernuzzi, S.
\newblock {GW170817: Joint Constraint on the Neutron Star Equation of State
  from Multimessenger Observations}.
\newblock {\em Astrophys. J. Lett.} {\bf 2018}, {\em 852},~L29,
  \href{http://xxx.lanl.gov/abs/1711.03647}{{\normalfont
  [arXiv:astro-ph.HE/1711.03647]}}.
\newblock {\url{https://doi.org/10.3847/2041-8213/aaa402}}.

\bibitem[Kubis and Kutschera(1997)]{Kubis_1997}
Kubis, S.; Kutschera, M.
\newblock Nuclear matter in relativistic mean field theory with isovector
  scalar meson.
\newblock {\em Physics Letters B} {\bf 1997}, {\em 399},~191--195.
\newblock {\url{https://doi.org/10.1016/s0370-2693(97)00306-7}}.

\bibitem[Kubis et~al.(1998)Kubis, Kutschera, and
  Stachniewicz]{https://doi.org/10.48550/arxiv.astro-ph/9802303}
Kubis, S.; Kutschera, M.; Stachniewicz, S.
\newblock Neutron Stars in Relativistic Mean Field Theory with Isovector Scalar
  Meson {\bf 1998}.
\newblock {\url{https://doi.org/10.48550/ARXIV.ASTRO-PH/9802303}}.

\bibitem[Miyatsu et~al.(2022)Miyatsu, Cheoun, and Saito]{Miyatsu_2022}
Miyatsu, T.; Cheoun, M.K.; Saito, K.
\newblock Asymmetric Nuclear Matter in Relativistic Mean-field Models with
  Isoscalar- and Isovector-meson Mixing.
\newblock {\em The Astrophysical Journal} {\bf 2022}, {\em 929},~82.
\newblock {\url{https://doi.org/10.3847/1538-4357/ac5f40}}.

\bibitem[Li et~al.(2022)Li, Cai, Zhou, Jiang, and Chen]{Li_2022}
Li, F.; Cai, B.J.; Zhou, Y.; Jiang, W.Z.; Chen, L.W.
\newblock Effects of Isoscalar- and Isovector-scalar Meson Mixing on Neutron
  Star Structure.
\newblock {\em The Astrophysical Journal} {\bf 2022}, {\em 929},~183.
\newblock {\url{https://doi.org/10.3847/1538-4357/ac5e2a}}.

\bibitem[Miyatsu et~al.(2022)Miyatsu, Cheoun, Kim, and
  Saito]{https://doi.org/10.48550/arxiv.2209.02861}
Miyatsu, T.; Cheoun, M.K.; Kim, K.; Saito, K.
\newblock Massive neutron stars with small radii in relativistic mean-field
  models optimized to nuclear ground states,  2022.
\newblock {\url{https://doi.org/10.48550/ARXIV.2209.02861}}.

\bibitem[Thakur et~al.(2022)Thakur, Kumar, Kumar, Kumar, Kumar, Mondal,
  Agrawal, and Dhiman]{Thakur_2022}
Thakur, V.; Kumar, R.; Kumar, P.; Kumar, V.; Kumar, M.; Mondal, C.; Agrawal,
  B.K.; Dhiman, S.K.
\newblock Effects of an isovector scalar meson on the equation of state of
  dense matter within a relativistic mean field model.
\newblock {\em Physical Review C} {\bf 2022}, {\em 106}.
\newblock {\url{https://doi.org/10.1103/physrevc.106.045806}}.

\bibitem[Liu et~al.(2005)Liu, Guo, Toro, and Greco]{Liu_2005}
Liu, B.; Guo, H.; Toro, M.D.; Greco, V.
\newblock Neutron stars with isovector scalar correlations.
\newblock {\em The European Physical Journal A} {\bf 2005}, {\em 25},~293--298.
\newblock {\url{https://doi.org/10.1140/epja/i2005-10095-1}}.

\bibitem[Rabhi et~al.(2009)Rabhi, Provid\^encia, and
  Provid\^encia]{PhysRevC.80.025806}
Rabhi, A.; Provid\^encia, C.; Provid\^encia, J.D.
\newblock Effect of the \ensuremath{\delta} meson on the instabilities of
  nuclear matter under strong magnetic fields.
\newblock {\em Phys. Rev. C} {\bf 2009}, {\em 80},~025806.
\newblock {\url{https://doi.org/10.1103/PhysRevC.80.025806}}.

\bibitem[Gaitanos et~al.(2004)Gaitanos, Toro, Typel, Baran, Fuchs, Greco, and
  Wolter]{Gaitanos_2004}
Gaitanos, T.; Toro, M.D.; Typel, S.; Baran, V.; Fuchs, C.; Greco, V.; Wolter,
  H.
\newblock On the Lorentz structure of the symmetry energy.
\newblock {\em Nuclear Physics A} {\bf 2004}, {\em 732},~24--48.
\newblock {\url{https://doi.org/10.1016/j.nuclphysa.2003.12.001}}.

\bibitem[Greco et~al.(2003)Greco, Colonna, Di~Toro, and
  Matera]{PhysRevC.67.015203}
Greco, V.; Colonna, M.; Di~Toro, M.; Matera, F.
\newblock Collective modes of asymmetric nuclear matter in quantum
  hadrodynamics.
\newblock {\em Phys. Rev. C} {\bf 2003}, {\em 67},~015203.
\newblock {\url{https://doi.org/10.1103/PhysRevC.67.015203}}.

\bibitem[Liu et~al.(2002)Liu, Greco, Baran, Colonna, and
  Di~Toro]{PhysRevC.65.045201}
Liu, B.; Greco, V.; Baran, V.; Colonna, M.; Di~Toro, M.
\newblock Asymmetric nuclear matter: The role of the isovector scalar channel.
\newblock {\em Phys. Rev. C} {\bf 2002}, {\em 65},~045201.
\newblock {\url{https://doi.org/10.1103/PhysRevC.65.045201}}.

\bibitem[Wang et~al.(2014)Wang, Zhang, and Dong]{PhysRevC.90.055801}
Wang, S.; Zhang, H.F.; Dong, J.M.
\newblock Neutron star properties in density-dependent relativistic mean field
  theory with consideration of an isovector scalar meson.
\newblock {\em Phys. Rev. C} {\bf 2014}, {\em 90},~055801.
\newblock {\url{https://doi.org/10.1103/PhysRevC.90.055801}}.

\bibitem[Roca-Maza et~al.(2011)Roca-Maza, Vi\~nas, Centelles, Ring, and
  Schuck]{PhysRevC.84.054309}
Roca-Maza, X.; Vi\~nas, X.; Centelles, M.; Ring, P.; Schuck, P.
\newblock Relativistic mean-field interaction with density-dependent
  meson-nucleon vertices based on microscopical calculations.
\newblock {\em Phys. Rev. C} {\bf 2011}, {\em 84},~054309.
\newblock {\url{https://doi.org/10.1103/PhysRevC.84.054309}}.

\bibitem[Kong et~al.(2023)Kong, Minamikawa, and Harada]{PhysRevC.108.055206}
Kong, Y.K.; Minamikawa, T.; Harada, M.
\newblock Neutron star matter based on a parity doublet model including the
  ${a}_{0}(980)$ meson.
\newblock {\em Phys. Rev. C} {\bf 2023}, {\em 108},~055206.
\newblock {\url{https://doi.org/10.1103/PhysRevC.108.055206}}.

\bibitem[Mun et~al.(2023)Mun, Shin, Paeng, Harada, and Kim]{mun2023}
Mun, M.H.; Shin, I.J.; Paeng, W.G.; Harada, M.; Kim, Y.
\newblock Nuclear structure in parity doublet model.
\newblock {\em The European Physical Journal A} {\bf 2023}, {\em 59},~149.
\newblock {\url{https://doi.org/10.1140/epja/s10050-023-01064-x}}.

\bibitem[Meng et~al.(2006)Meng, Toki, Zhou, Zhang, Long, and Geng]{Meng:2005jv}
Meng, J.; Toki, H.; Zhou, S.G.; Zhang, S.Q.; Long, W.H.; Geng, L.S.
\newblock {Relativistic Continuum Hartree Bogoliubov theory for ground state
  properties of exotic nuclei}.
\newblock {\em Prog. Part. Nucl. Phys.} {\bf 2006}, {\em 57},~470--563,
  \href{http://xxx.lanl.gov/abs/nucl-th/0508020}{{\normalfont
  [nucl-th/0508020]}}.
\newblock {\url{https://doi.org/10.1016/j.ppnp.2005.06.001}}.

\bibitem[Bando et~al.(1985)Bando, Kugo, Uehara, Yamawaki, and
  Yanagida]{PhysRevLett.54.1215}
Bando, M.; Kugo, T.; Uehara, S.; Yamawaki, K.; Yanagida, T.
\newblock Is the $\ensuremath{\rho}$ Meson a Dynamical Gauge Boson of Hidden
  Local Symmetry?
\newblock {\em Phys. Rev. Lett.} {\bf 1985}, {\em 54},~1215--1218.
\newblock {\url{https://doi.org/10.1103/PhysRevLett.54.1215}}.

\bibitem[Bando et~al.(1988)Bando, Kugo, and Yamawaki]{Bando:1987br}
Bando, M.; Kugo, T.; Yamawaki, K.
\newblock {Nonlinear Realization and Hidden Local Symmetries}.
\newblock {\em Phys. Rept.} {\bf 1988}, {\em 164},~217--314.
\newblock {\url{https://doi.org/10.1016/0370-1573(88)90019-1}}.

\bibitem[Harada and Yamawaki(2003)]{Harada:2003jx}
Harada, M.; Yamawaki, K.
\newblock {Hidden local symmetry at loop: A New perspective of composite gauge
  boson and chiral phase transition}.
\newblock {\em Phys. Rept.} {\bf 2003}, {\em 381},~1--233,
  \href{http://xxx.lanl.gov/abs/hep-ph/0302103}{{\normalfont
  [hep-ph/0302103]}}.
\newblock {\url{https://doi.org/10.1016/S0370-1573(03)00139-X}}.

\bibitem[Garg and Col{\`o}(2018)]{GARG201855}
Garg, U.; Col{\`o}, G.
\newblock The compression-mode giant resonances and nuclear incompressibility.
\newblock {\em Progress in Particle and Nuclear Physics} {\bf 2018}, {\em
  101},~55--95.
\newblock {\url{https://doi.org/https://doi.org/10.1016/j.ppnp.2018.03.001}}.

\bibitem[Li et~al.(2021)Li, Cai, Xie, and Zhang]{universe7060182}
Li, B.A.; Cai, B.J.; Xie, W.J.; Zhang, N.B.
\newblock Progress in Constraining Nuclear Symmetry Energy Using Neutron Star
  Observables Since GW170817.
\newblock {\em Universe} {\bf 2021}, {\em 7}.
\newblock {\url{https://doi.org/10.3390/universe7060182}}.

\bibitem[Sunardi and Sulaksono(2018)]{Sunardi:2018snn}
Sunardi, A.; Sulaksono, A.
\newblock {Nuclear bubble in Sn isotope within modified relativistic mean
  field}.
\newblock {\em AIP Conf. Proc.} {\bf 2018}, {\em 2023},~020007.
\newblock {\url{https://doi.org/10.1063/1.5064004}}.

\end{thebibliography}

\end{adjustwidth}
\end{document}